\documentclass[11pt,a4paper]{article}
\usepackage{amsmath, amsfonts, amssymb}
\usepackage{a4wide}
\usepackage{parskip}
\usepackage{enumitem}
\usepackage{xcolor}

\usepackage{hyperref}
\usepackage{booktabs}
\usepackage{caption}
\usepackage{siunitx}
\usepackage{tabularx}
\usepackage{graphicx}
\usepackage{adjustbox}
\usepackage{multirow}
\usepackage{amsthm}
\usepackage{bm}
\usepackage{lscape}
\usepackage{float}
\usepackage{comment}
\usepackage[noadjust]{cite}

\newtheorem{theorem}{Theorem}[section]
\newtheorem{lemma}[theorem]{Lemma}
\theoremstyle{definition}
\newtheorem{definition}[theorem]{Definition}
\newtheorem{example}[theorem]{Example}
\newtheorem{proposition}[theorem]{Proposition}
\theoremstyle{remark}
\newtheorem{remark}[theorem]{Remark}

\newtheorem{corollary}[theorem]{Corollary}

\numberwithin{equation}{section}

\newcommand{\Char}{\textnormal{char}}

\newcommand{\Span}{\textnormal{Span}}

\usepackage{tikz,xcolor,hyperref}

\definecolor{lime}{HTML}{A6CE39}
\DeclareRobustCommand{\orcidicon}{%
	\begin{tikzpicture}
		\draw[lime, fill=lime] (0,0) 
		circle [radius=0.16] 
		node[white] {{\fontfamily{qag}\selectfont \tiny ID}};
		\draw[white, fill=white] (-0.0625,0.095) 
		circle [radius=0.007];
	\end{tikzpicture}
	\hspace{-2mm}
}

\foreach \x in {A, ..., Z}{%
	\expandafter\xdef\csname orcid\x\endcsname{\noexpand\href{https://orcid.org/\csname orcidauthor\x\endcsname}{\noexpand\orcidicon}}
}

\begin{document}
	\date{}
	{\vspace{0.01in}
		\title{Non-RS MDS Codes via Row-Column Twists}

        \author{{\bf Anuj Kumar Bhagat\footnote{email: {\tt anujkumarbhagat632@gmail.com}}\orcidA{},\;\bf Harshdeep Singh\footnote{email: {\tt harshdeep.sag@gov.in}}\orcidH{}, and \bf Ritumoni Sarma\footnote{    email: {\tt ritumoni407@gmail.com}}\orcidR{}} \\ $^{\ast \dagger\ddagger}$Department of Mathematics,\\ Indian Institute of Technology Delhi,\\Hauz Khas, New Delhi-110016, India \medskip \\}
\maketitle
\begin{abstract}
    This article introduces a new class of codes, called row-column twisted Reed--Solomon (RCTRS) codes, motivated by the constructions in (Beelen et al. \cite{beelen2017twisted}) and (Liu et al. \cite{liu2025column}). Explicit conditions under which RCTRS codes are MDS are established, and their existence is proved by algebraic techniques. By deriving lower bounds on the dimensions of their Schur squares, it is shown that these MDS codes, as well as their extended codes, are not equivalent to Reed--Solomon codes, and hence form new families of non-RS MDS codes. It is further proved that they are not equivalent to column twisted Reed--Solomon codes or their extended versions, and, when the hook lies strictly between the two extreme positions, are not equivalent to twisted Reed--Solomon codes with twist $t = 1$ either. Introducing twists in both a row and a column therefore yields a family of MDS codes distinct from the previously known constructions.


    \medskip
		\noindent \textit{Keywords:}
			Generalized Reed--Solomon (GRS) codes, Twisted Reed--Solomon (TRS) codes, Column twisted Reed--Solomon (CTRS) codes, Non-RS MDS codes, Schur square.
			\medskip

            \noindent\textit{MSC2020:}  94B65, 05B25
\end{abstract}
\section{Introduction}\label{Section 1}
    An $[n, k, d]$-linear code over a finite field $\mathbb{F}_q$ is a $k$-dimensional subspace of the $\mathbb{F}_q$-vector space $\mathbb{F}_q^n$ with Hamming distance $d.$ For any $[n,k,d]$-linear code, the Singleton bound states that $d\le n-k+1;$ and linear codes that meet this bound are called maximal distance separable (in short, MDS) codes. Such codes maximize the error-detection and error-correction capabilities of the code by achieving the maximum possible Hamming distance for the specified length and dimension. Consequently, there has been significant interest among researchers in various aspects of MDS codes, such as their classification up to equivalence \cite{kokkala2015classification}, their weight distribution \cite{alderson2020weights}, \cite{ezerman2010weights}, their covering radius \cite{bartoli2014covering}, linear complementary dual (LCD) properties \cite{carlet2018euclidean}, as well as their applications in cryptography \cite{Niederreiter1986Knapsack}. Moreover, MDS codes find applications in combinatorial designs and are closely connected to finite geometry \cite{hirschfeld1998projective}, \cite{macwilliams1977theory}.

    A well-known and important family of maximum distance separable (MDS) codes over finite fields is the family of Generalized Reed-Solomon (GRS) codes \cite{Roth_2006}, whose length is at most equal to the size of the underlying field. More precisely, extended GRS codes over $\mathbb{F}_q$ are also MDS, with their length bounded above by $q+1$. Moreover, the dual of a GRS code is also a GRS code. There is a well-known conjecture on MDS codes which states that the length of an MDS code over $\mathbb{F}_q$ cannot exceed $q + 2$, and, more specifically, is at most $q + 1$ except for a few cases \cite{segre1955curve}. Significant progress has been made toward proving this conjecture; notably, it has been established that the conjecture holds for prime fields \cite{ball2012large}. To date, the well-known MDS codes are GRS and extended GRS codes, and most known MDS codes are equivalent to one of them. 
    Finding new MDS codes that are not equivalent to GRS codes has always been a challenging problem. 
    It is a well-established fact that any $[n, k]$-MDS code is equivalent to a Reed-Solomon code when $k<3$ or $n-k<3$ (\cite{beelen2017twisted}, Corr. $2$). 
    Therefore, non-RS equivalent $[n,k]$-MDS codes are restricted to $3 \leq k\leq n-3$. In addition, since the dual of a GRS code is again a GRS code, it suffices to focus on the case where $3 \leq k \leq n/2$ in the subsequent discussion.

    An MDS code that is not equivalent to a GRS code or an extended GRS code is referred to as a non-RS MDS code. The first such construction was given in 1989 by Roth and Lempel in \cite{RothLempel1989}, based on generator matrices and special subsets of finite fields.  More recently, in 2016, Sheekey in \cite{JohnSheekey2016} introduced a new class of maximum rank distance codes, known as Twisted Gabidulin codes, which are MDS with respect to the rank metric and were shown to be inequivalent to Gabidulin codes (the rank-metric counterpart of Reed-Solomon codes). Inspired by the work of Sheekey, Beelen et al. in \cite{beelen2017twisted} and  \cite{beelen2022twisted} proposed Twisted Reed-Solomon (TRS) codes and demonstrated that certain subfamilies of TRS are non-RS MDS codes. Following the work of Beelen et al., many studies have focused on the structure and properties of TRS codes, including their duality, see for instance,  \cite{Beelen_Structural_2018}, \cite{cheng2023parity}, \cite{huang2021mds}, \cite{sui2022mds}, \cite{zhang2022class}, \cite{sharma2018multi} and  \cite{singh2024mds}. Furthermore, many families of non-RS MDS codes have been constructed using different techniques, see for instance, \cite{abdukhalikov2025some}, \cite{zhi2025new}, \cite{jin2024new}, \cite{li2025non}, \cite{wu2024more} and \cite{chen2023many}.

    Recently, Liu et al. \cite{liu2025column} introduced column-twisted Reed--Solomon (CTRS) codes by adding a twist to a column of the generator matrix of an RS code. They established conditions under which these codes are MDS and, by analyzing the dimension of their Schur squares, showed that CTRS codes yield non-RS MDS codes. Motivated by the constructions of Beelen et al. \cite{beelen2017twisted} and Liu et al. \cite{liu2025column}, we introduce a new family of codes, called row-column twisted Reed--Solomon (RCTRS) codes, obtained by introducing twists in both a row and a column of the generator matrix. We derive necessary and sufficient conditions for these codes to be MDS and construct explicit families satisfying these conditions. By establishing lower bounds on the dimensions of their Schur squares, we show that these codes are non-RS MDS codes and are not equivalent to either column-twisted Reed--Solomon codes or their extended versions. Moreover, when the hook lies strictly between its two extreme positions, the bound improves by one, which also rules out equivalence with twisted Reed--Solomon codes with twist $t=1$.

    The rest of the article is organized as follows. Section \ref{Section 2} presents the preliminaries, including standard material on GRS codes, Schur products, and code equivalence, together with the two key observations underlying our Schur square computations: Lemma \ref{lem:: schur square upper bound}, which bounds the Schur square dimension of an evaluation code with finitely many additional coordinates, and Lemma \ref{lem:: puncturing}, which shows that puncturing cannot increase the Schur square dimension. Section \ref{Section 3} introduces RCTRS codes and their extended codes, gives their generator matrices, and presents results on the Schur squares of CTRS, TRS, and RCTRS codes. The next three sections treat the three possible positions of the hook. Section \ref{Section 4} considers $(h,t)=(0,1)$ and establishes MDS conditions, constructions, and inequivalence results for the resulting RCTRS codes and their extended versions. Section \ref{Section 5} follows the same programme for $(h,t)=(k-1,1)$. Section \ref{Section 6} treats the general case $1\le h\le k-2$: we give MDS conditions and constructions and prove that the resulting codes and their extended versions are inequivalent to any RS or CTRS codes, their extended versions, or any TRS code with twist $t=1$. Table \ref{tab:: comparison} summarises the main inequivalence results, and Section \ref{Section 7} concludes the article.

\section{Preliminaries}\label{Section 2}
Throughout this article, for a prime power $q,$ the field of order $q$ is denoted by $\mathbb{F}_q$. Let $\mathbb{F}_q[x]_{<k}:=\{f(x)\in\mathbb{F}_q[x]: \deg f(x)< k\}.$
\begin{definition}\cite{Roth_2006}
    For $k\le n\le q,$ let $\bm{\alpha}:=(\alpha_1,\alpha_2, \dots, \alpha_n)\in\mathbb{F}_q^n$ with $\alpha_i\ne \alpha_j,$ for $i\ne j$ and $\bm{v}:=(v_1, v_2, \dots, v_n)\in(\mathbb{F}_q^*)^n$. A \textit{Generalized Reed-Solomon (GRS) code} is defined as
    \begin{equation}
        \textnormal{GRS}_{k,n}(\bm{\alpha}, \bm{v}):=\{(v_1f(\alpha_1), v_2f(\alpha_2), \dots, v_nf(\alpha_n)): f(x)\in\mathbb{F}_q[x]_{<k}\},
    \end{equation}
    and an \textit{extended GRS code} is defined as
    \begin{equation}
        \textnormal{GRS}_{k,n}(\bm{\alpha}, \bm{v}, \infty):=\{(v_1f(\alpha_1), v_2f(\alpha_2), \dots, v_nf(\alpha_n), f_{k-1}): f(x)\in\mathbb{F}_q[x]_{<k}\},
    \end{equation}
    where $f_{k-1}$ is the coefficient of $x^{k-1}$ in $f(x).$
\end{definition}
If $\bm{v}=\bm{1}:=(1, 1, \dots, 1),$ then the corresponding GRS and the extended GRS codes are called \textit{Reed-Solomon (RS)} and \textit{extended RS codes}, respectively.

The parameters of the codes $\textnormal{GRS}_{k,n}(\bm{\alpha}, \bm{v})$ and $\textnormal{GRS}_{k,n}(\bm{\alpha}, \bm{v}, \infty)$ are $[n, k, n-k+1]$ and $[n+1, k, n-k+2],$ respectively; hence, both of these codes are MDS.
\begin{definition}\cite{beelen2017twisted}
    For $k, t\in \mathbb{N}$ and $h\in \mathbb{N}\cup \{0\}$, satisfying $0\le h<k\le q,$ define 
    \begin{equation}
        \mathcal{V}_{k, h, t, \eta}:=\left\{f(x)=\sum_{i=0}^{k-1} f_i x^{i}+\eta f_h x^{k-1+t}: f_i \in\mathbb{F}_q\right\},
    \end{equation}
    for $\eta\in\mathbb{F}_q.$ For $\mathcal{V}_{k, h, t, \eta},$ the parameters $h$ and $t$ are called hook and twist, respectively. 
\end{definition}
\begin{definition}\cite{beelen2017twisted}\label{defn:: TRS}
    For $k\le n\le q,$ let $\bm{\alpha}:=(\alpha_1,\alpha_2, \dots, \alpha_n)\in\mathbb{F}_q^n,$ with $\alpha_i\ne \alpha_j,$ for $i\ne j.$ Suppose $\eta\in\mathbb{F}_q.$ A \textit{twisted Reed-Solomon (TRS) code} is given by
    \begin{equation}
        \textnormal{TRS}_{k,h,t}(\bm{\alpha}, \eta):=\{(f(\alpha_1), f(\alpha_2), \dots, f(\alpha_n)): f(x)\in\mathcal{V}_{k, h, t, \eta}\}.
    \end{equation}
\end{definition}
\begin{definition}
    The Schur product of $\bm{x}=(x_1, x_2, \dots, x_n)\in\mathbb{F}_q^n$ and $\bm{y}=(y_1, y_2, \dots, y_n)\in\mathbb{F}_q^n$ is given by $\bm{x}\star \bm{y}:=(x_1y_1, x_2y_2, \dots, x_ny_n).$ For two linear codes $\mathcal{C}$ and $\mathcal{D}$ over the same field and of the same length, their \textit{Schur product} is given by
    \begin{equation}
        \mathcal{C}\star\mathcal{D}:=\Span_{\mathbb{F}_q}\{\bm{c}\star\bm{d}: \bm{c}\in\mathcal{C},\bm{d}\in\mathcal{D}\}.
    \end{equation}
\end{definition}
In particular, $\mathcal{C}^{\star 2},$ the Schur product of $\mathcal{C}$ with itself is often called the \textit{Schur square} of $\mathcal{C}.$

Now, we record a lemma on the Schur square of a GRS code and an extended GRS code.
\begin{lemma}\cite{couvreur2014distinguisher}\label{lem:: schur square of GRS}
    For $k\le \frac{n}{2},$ $\textnormal{GRS}_{k,n}(\bm{\alpha}, \bm{v})^{\star 2}=\textnormal{GRS}_{2k-1, n}(\bm{\alpha}, \bm{v}\star\bm{v})$ and $\textnormal{GRS}_{k,n-1}(\bm{\beta}, \bm{v}, \infty)^{\star 2}\\=\textnormal{GRS}_{2k-1,n-1}(\bm{\beta}, \bm{v}\star\bm{v}, \infty)$, where $\bm\alpha\in\mathbb{F}_q^n$ and $\bm\beta\in\mathbb{F}_q^{n-1}$ with both having distinct coordinates.
\end{lemma}
\begin{definition}\cite{beelen2017twisted}
    Two $[n,k]$-codes $\mathcal{C}$ and  $\mathcal{D}$ over $\mathbb{F}_q$ are called \textit{equivalent} if there is a permutation $\sigma\in S_n$ and $\bm{v}=(v_1, v_2, \dots, v_n)\in(\mathbb{F}_q^{*})^n$ such that $\mathcal{C}=\psi_{\sigma, \bm{v}}(\mathcal{D}),$ where 
    \begin{align*}
    \psi_{\sigma, \bm{v}}:\mathbb{F}_q^n&\to\mathbb{F}_q^n\\
    (x_1, x_2, \dots, x_n)&\mapsto (v_1 x_{\sigma(1)}, v_2 x_{\sigma(2)}, \dots, v_n x_{\sigma(n)})
    \end{align*}
    is the Hamming-metric isometry of $\mathbb{F}_q^n.$
\end{definition}
\begin{remark}\label{remark:: non RS MDS code}
    If two linear codes are equivalent, then their Schur squares are also equivalent. Any MDS code that is equivalent to neither a GRS nor an extended GRS code is referred to as a \textit{non-RS MDS code} in this article. Hence, if $\mathcal{C}$ is an $[n,k, n-k+1]$-code over $\mathbb{F}_q$ with $k\le \frac{n}{2}$ such that $\dim \mathcal{C}^{\star 2}\ne 2k-1,$ then $\mathcal{C}$ is a non-RS MDS code.
\end{remark}
All the codes considered in this article are obtained by evaluating polynomials at a prescribed set of points and appending a bounded number of additional coordinates. For such codes, the dimension of the Schur square is governed by a purely polynomial quantity, as formalized in the following lemma.

For an $\mathbb{F}_q$-subspace $V\subseteq \mathbb{F}_q[x]$, we write
$$V^{\star 2}:=\Span_{\mathbb{F}_q}\{fg: f, g\in V\}\subseteq \mathbb{F}_q[x].$$

\begin{lemma}\label{lem:: schur square upper bound}
    Let $\alpha_1, \alpha_2, \dots, \alpha_m\in\mathbb{F}_q$ be pairwise distinct, let $V\subseteq\mathbb{F}_q[x]$ be an $\mathbb{F}_q$-subspace, and let $L_1, L_2, \dots, L_r: V\to \mathbb{F}_q$ be $\mathbb{F}_q$-linear maps. If
    $$\mathcal{C}:=\{(f(\alpha_1), \dots, f(\alpha_m), L_1(f), \dots, L_r(f)): f\in V\}\subseteq \mathbb{F}_q^{m+r},$$
    then $\dim (\mathcal{C}^{\star 2})\le \dim (V^{\star 2})+r.$
\end{lemma}
\begin{proof}
    Let
    $$\mathcal{W}:=\{(P(\alpha_1), \dots, P(\alpha_m), y_1, \dots, y_r): P\in V^{\star 2},\ y_1, \dots, y_r\in\mathbb{F}_q\}.$$
    Then $\mathcal{W}$ is the image of the $\mathbb{F}_q$-linear map $V^{\star 2}\oplus \mathbb{F}_q^{r}\to \mathbb{F}_q^{m+r}$ which sends $(P, y_1, \dots, y_r)$ to $(P(\alpha_1), \dots, P(\alpha_m), y_1, \dots, y_r),$ so that $\dim\mathcal{W}\le \dim (V^{\star 2})+r.$ For $f, g\in V,$ the Schur product of the two corresponding codewords of $\mathcal{C}$ is
    $$\bigl((fg)(\alpha_1), \dots, (fg)(\alpha_m), L_1(f)L_1(g), \dots, L_r(f)L_r(g)\bigr).$$
    Since $fg\in V^{\star 2},$ this vector belong to $\mathcal{W}.$ Such Schur products span $\mathcal{C}^{\star 2},$ and therefore $\mathcal{C}^{\star 2}\subseteq \mathcal{W}.$ \hfill$\square$
\end{proof}

\begin{lemma}\label{lem:: puncturing}
    Let $\mathcal{D}\subseteq \mathbb{F}_q^{N+1}$ be a linear code and let $\mathcal{C}\subseteq\mathbb{F}_q^{N}$ be the code obtained from $\mathcal{D}$ by deleting the last coordinate. Then $\dim (\mathcal{D}^{\star 2})\ge \dim (\mathcal{C}^{\star 2}).$
\end{lemma}
\begin{proof}
    Let $\pi:\mathbb{F}_q^{N+1}\to\mathbb{F}_q^{N}$ delete the last coordinate, so that $\pi(\mathcal{D})=\mathcal{C}.$ Then $\pi(\bm x\star \bm y)=\pi(\bm x)\star\pi(\bm y)$ for all $\bm x, \bm y\in \mathbb{F}_q^{N+1}.$ Hence $\pi$ carries the Schur products $\bm d_1\star \bm d_2$ with $\bm d_1, \bm d_2\in\mathcal{D}$ onto the Schur products $\bm c_1\star \bm c_2$ with $\bm c_1, \bm c_2\in\mathcal{C},$ and since these span $\mathcal{D}^{\star 2}$ and $\mathcal{C}^{\star 2}$ respectively, $\pi(\mathcal{D}^{\star 2})=\mathcal{C}^{\star 2}.$ Since the image of a linear map cannot have larger dimension than its domain, $\dim(\mathcal{D}^{\star 2})\ge \dim(\mathcal{C}^{\star 2}).$ \hfill$\square$
\end{proof}
The following lemma is useful to determine if a code is MDS.
\begin{lemma}\cite{macwilliams1977theory}
     An $[n,k]$-linear code over $\mathbb{F}_q$ is MDS if and only if each set of $k$ distinct columns of its generator matrix is linearly independent.
\end{lemma}
\begin{definition}
    For $\alpha_1, \alpha_2, \dots, \alpha_n\in \mathbb{F}_q, $ the matrix
    \begin{equation}
        V(\alpha_1, \dots, \alpha_n):=\begin{bmatrix}
            1 & 1 & \dots & 1\\
            \alpha_1 & \alpha_2 & \dots &\alpha_n\\
            \alpha_1^2 & \alpha_2^2 & \dots &\alpha_n^2\\
            \vdots & \vdots & \dots & \vdots\\
            \alpha_1^{n-1} & \alpha_2^{n-1} & \dots &\alpha_n^{n-1}\\
        \end{bmatrix}
    \end{equation}
    is called a \textit{Vandermonde matrix} of order $n$.
\end{definition}
It is well-known that $\det V(\alpha_1, \dots, \alpha_n)=\underset{1\le i<j\le n}{\prod}(\alpha_j-\alpha_i).$
\begin{definition}\cite{Lang2002Algebra}
    For $r\in \mathbb{N},$ the $r$th \textit{elementary symmetric function} in $N$ variables $X_1, X_2$ $\dots, X_N$ is defined as
    \begin{equation}
        \sigma_{r}(X_1, X_2, \dots, X_N):=\underset{1\le i_1< i_2< \dots < i_r\le N}{\sum} X_{i_1} X_{i_2}\cdots X_{i_r},
    \end{equation}
    for $0<r\le N$ and $\sigma_r(X_1, X_2, \dots, X_N)=0,$ for $r>N.$ Sometimes, it is also conventional to define $\sigma_0(X_1, X_2, \dots, X_N)=1.$
\end{definition}
\begin{proposition}\cite{muir2003treatise}\label{prop:: Det of vandermonde with a row deleted}
    For $1\le h\le n-1,$ the determinant of the matrix
    \begin{equation}
        \begin{bmatrix}
            1 & 1 & \dots & 1\\
            \alpha_1 & \alpha_2 & \dots &\alpha_n\\
            \vdots & \vdots & \dots & \vdots\\
            \alpha_1^{h-1} & \alpha_2^{h-1} & \dots &\alpha_n^{h-1}\\
            \alpha_1^{h+1} & \alpha_2^{h+1} & \dots &\alpha_n^{h+1}\\
            \vdots & \vdots & \dots & \vdots\\
            \alpha_1^{n} & \alpha_2^{n} & \dots &\alpha_n^{n}\\
        \end{bmatrix}
    \end{equation}
     is $\sigma_{n-h}(\alpha_1, \alpha_2, \dots, \alpha_n) \det V(\alpha_1, \alpha_2, \dots, \alpha_n).$
\end{proposition}
\section{Row-Column Twisted Reed-Solomon Codes}\label{Section 3}
    In this short section, we first recall the notion of column-twisted Reed--Solomon (CTRS) codes introduced in \cite{liu2025column}. Motivated by \cite{beelen2017twisted} and \cite{liu2025column}, we then define row-column twisted Reed--Solomon (RCTRS) codes and their extended versions as natural generalizations of the constructions in these works. We also determine their generator matrices and establish results concerning the Schur squares of CTRS, TRS, and RCTRS codes, as well as their extended versions.
    
    In \cite{liu2025column}, column-twisted Reed-Solomon codes are introduced through their generator matrices. Nevertheless, the same construction can be reformulated in an algebraic manner, as described below.
    \begin{definition}\label{defn:: CTRS}
    For $k\le n\le q+1,$ let $\bm{\alpha}:=(\alpha_1,\alpha_2, \dots, \alpha_{n-1})\in\mathbb{F}_q^{n-1},$ where $\alpha_i\ne \alpha_j$, for $i\ne j$ and let
    $b,c, \lambda\in \mathbb{F}_q.$ A \textit{column-twisted Reed-Solomon (CTRS) code} is given by
    \begin{equation}
        \textnormal{CTRS}(\bm{\alpha}, b, c, \lambda):=\{(f(\alpha_1), f(\alpha_2), \dots, f(\alpha_{n-1}), f(b)-\lambda f(c)): f(x)\in\mathbb{F}_q[x]_{<k}\},
    \end{equation}
    and an \textit{extended CTRS code} is given by
    \begin{equation}
        \textnormal{CTRS}(\bm{\alpha}, b,c,\lambda, \infty):=\{(f(\alpha_1), f(\alpha_2), \dots, f(\alpha_{n-1}), f(b)-\lambda f(c), f_{k-1}): f(x)\in\mathbb{F}_q[x]_{<k}\},
    \end{equation}
    where $f_{k-1}$ is the coefficient of $x^{k-1}$ in $f(x).$
\end{definition}

\begin{lemma}\label{cor:: schur square bounds}
    Let $k\le n\le q+1$ and let $\bm\alpha\in\mathbb{F}_q^{n-1}$ have pairwise distinct coordinates.
    \begin{enumerate}
        \item [(a)] For all $b, c, \lambda\in\mathbb{F}_q,$ $\dim \textnormal{CTRS}(\bm\alpha, b, c, \lambda)^{\star 2}\le 2k$ and $\dim \textnormal{CTRS}(\bm\alpha, b, c, \lambda, \infty)^{\star 2}\le 2k.$
        \item [(b)] For all $\bm\beta\in\mathbb{F}_q^{n}$ with pairwise distinct coordinates, all $\eta\in\mathbb{F}_q$ and all $h$ with $0\le h\le k-1,$ $\dim \textnormal{TRS}_{k, h, 1}(\bm\beta, \eta)^{\star 2}\le 2k+1.$
        \item [(c)] For all $\bm\beta\in\mathbb{F}_q^{n}$ with pairwise distinct coordinates and all $\eta\in\mathbb{F}_q,$ $\dim \textnormal{TRS}_{k, 0, 1}(\bm\beta, \eta)^{\star 2}\le 2k.$
    \end{enumerate}
\end{lemma}
\begin{proof}
    (a) Take $V=\mathbb{F}_q[x]_{<k},$ $m=n-1$ and $r=1$ with $L_1(f)=f(b)-\lambda f(c).$ Every product of two polynomials of degree less than $k$ has degree at most $2k-2,$ so $V^{\star 2}=\mathbb{F}_q[x]_{<2k-1}$ and $\dim V^{\star 2}=2k-1;$ Lemma \ref{lem:: schur square upper bound} gives the first bound. For the extended code, observe that for $f, g\in\mathbb{F}_q[x]_{<k}$ the product $f_{k-1}g_{k-1}$ of the last coordinates is the coefficient of $x^{2k-2}$ in $fg.$ Hence every Schur product of two codewords lies in
    $$\{(P(\alpha_1), \dots, P(\alpha_{n-1}), y, P_{2k-2}): P\in\mathbb{F}_q[x]_{<2k-1},\ y\in\mathbb{F}_q\},$$
    where $P_{2k-2}$ denotes the coefficient of $x^{2k-2}$ in $P.$ This set is the image of an $\mathbb{F}_q$-linear map defined on $\mathbb{F}_q[x]_{<2k-1}\oplus\mathbb{F}_q,$ so its dimension is at most $2k,$ and the second bound follows.

    (b) Taking $V=\mathcal{V}_{k, h, 1, \eta},$ $m=n$, and $r=0,$ every polynomial in $\mathcal{V}_{k, h, 1, \eta}$ has degree at most $k$. Consequently, the product of any two such polynomials has degree at most $2k$, and hence $V^{\star 2}\subseteq \mathbb{F}_q[x]_{\le 2k}.$ It follows that $\dim V^{\star 2}\le 2k+1.$ Applying Lemma \ref{lem:: schur square upper bound} gives the required bound.

    (c) Let $V=\mathcal{V}_{k, 0, 1, \eta},$ which has basis $\{1+\eta x^{k}, x, x^2, \dots, x^{k-1}\}.$ Write $P_j$ for the coefficient of $x^{j}$ in $P\in\mathbb{F}_q[x]_{\le 2k},$ and let $\varphi:\mathbb{F}_q[x]_{\le 2k}\to\mathbb{F}_q$ be the $\mathbb{F}_q$-linear map $\varphi(P)=\eta^2P_0-P_{2k}.$ The space $V^{\star 2}$ is spanned by the products of pairs of basis elements, and each of these lies in $\ker\varphi$: the product $x^{i}x^{j}$ with $1\le i, j\le k-1,$ whose degree lies between $2$ and $2k-2,$ and hence its constant and $x^{2k}$-coefficients are both zero. Similarly, the product $(1+\eta x^{k})x^{i}=x^{i}+\eta x^{k+i}$ with $1\le i\le k-1$ have all their exponents between $1$ and $2k-1;$ and $\varphi\bigl((1+\eta x^{k})^2\bigr)=\varphi(1+2\eta x^{k}+\eta^2x^{2k})=\eta^2-\eta^2=0.$ Hence $V^{\star 2}\subseteq\ker\varphi.$ If $\eta\ne 0,$ then $\varphi$ is a nonzero linear form and $\dim\ker\varphi=2k,$ so $\dim V^{\star 2}\le 2k;$ if $\eta=0,$ then $V^{\star 2}=\mathbb{F}_q[x]_{<2k-1}$ has dimension $2k-1.$ In either case $\dim V^{\star 2}\le 2k,$ and Lemma \ref{lem:: schur square upper bound} with $r=0$ gives the required bound. \hfill$\square$
\end{proof}

\begin{definition}\label{defn:: RCTRS}
    For $0\le h< k\le n\le q+1,$ let $\bm{\alpha}:=(\alpha_1,\alpha_2, \dots, \alpha_{n-1})\in\mathbb{F}_q^{n-1},$ where $\alpha_i\ne \alpha_j$, for $i\ne j.$ 
    Suppose that $b,c, \lambda, \eta\in \mathbb{F}_q$ and $t\in\mathbb{N}.$ A \textit{row-column twisted Reed-Solomon (\textnormal{RCTRS}) code} is given by
    \begin{equation}
        \textnormal{RCTRS}_{h, t}(\bm{\alpha}, b, c, \lambda, \eta):=\{(f(\alpha_1), f(\alpha_2), \dots, f(\alpha_{n-1}), f(b)-\lambda f(c)): f(x)\in\mathcal{V}_{k, h, t, \eta}\},
    \end{equation}
    and an \textit{extended \textnormal{RCTRS} code} is given by
    \begin{equation}
        \textnormal{RCTRS}_{h,t}(\bm{\alpha}, b,c,\lambda, \eta, \infty):=\{(f(\alpha_1), \dots, f(\alpha_{n-1}), f(b)-\lambda f(c), f_{k-1}): f(x)\in\mathcal{V}_{k, h, t, \eta}\},
    \end{equation}
    where $f_{k-1}$ is the coefficient of $x^{k-1}$ in $f(x).$
\end{definition}
\begin{remark}
    If $b\ne \alpha_i$ for all $i\in\{1,\dots,n-1\},$ then   $\textnormal{RCTRS}_{h,t}(\bm{\alpha}, b,c,0, \eta)$ is $\textnormal{TRS}_{k,h,t}(\bm{\alpha}_b, \eta),$ where $\bm{\alpha}_b=(\alpha_1, \dots, \alpha_{n-1}, b)$ and $\textnormal{RCTRS}_{h,t}(\bm{\alpha}, b,c,\lambda, 0)$ is $\textnormal{CTRS}(\bm{\alpha}, b, c, \lambda).$ \\Moreover, $\textnormal{RCTRS}_{h,t}(\bm{\alpha}, b,$
    $c,0, 0)$ is a GRS$(\bm \alpha_b, \bm{1})$ code and $\textnormal{RCTRS}_{h,t}(\bm{\alpha}, b,c,0, 0, \infty)$ is an extended GRS$(\bm \alpha_b,\bm{1})$ code.
\end{remark}
\begin{theorem}
    The codes $\textnormal{RCTRS}_{h, t}(\bm{\alpha}, b, c, \lambda, \eta)$ and $\textnormal{RCTRS}_{h, t}(\bm{\alpha}, b, c, \lambda, \eta, \infty)$, as defined in Definition \ref{defn:: RCTRS}, have parameters $[n, k]$ and $[n+1, k]$ over $\mathbb{F}_q$ and their generator matrices are given in \eqref{generator matrix RCTRS} and \eqref{generator matrix RCTRS_inf}. 
\end{theorem}
\begin{proof}
    The length of these codes is clear from the definition. The set $\mathcal{B}:=\{1, x, x^2, \dots, x^{h-1},\\ x^h+\eta x^{k-1+t}, x^{h+1}, \dots, x^{k-1}\}$ is an $\mathbb{F}_q$-basis of $\mathcal{V}_{k, h, t, \eta}.$ It is easy to see that the set $$\{(f(\alpha_1), f(\alpha_2), \dots, f(\alpha_{n-1}), f(b)-\lambda f(c)): f(x)\in \mathcal{B}\}$$ is an $\mathbb{F}_q$-basis of $\textnormal{RCTRS}_{h, t}(\bm{\alpha}, b, c, \lambda, \eta)$ and the set $$\{(f(\alpha_1), f(\alpha_2), \dots, f(\alpha_{n-1}), f(b)-\lambda f(c), f_{k-1}): f(x)\in \mathcal{B}\}$$ is an $\mathbb{F}_q$-basis of $\textnormal{RCTRS}_{h, t}(\bm{\alpha}, b, c, \lambda, \eta, \infty).$ Since $|\mathcal{B}|=k,$ this proves that their dimensions are $k$. Consequently, a generator matrix of $\textnormal{RCTRS}_{h, t}(\bm{\alpha}, b, c, \lambda, \eta)$ is given by:
    {\small{
    \begin{equation}\label{generator matrix RCTRS}
        G_{h, t}(\bm \alpha,b,c,\lambda, \eta)=\begin{bmatrix}
            1 & \cdots & 1 & 1-\lambda\\
            \alpha_{1}  & \cdots & \alpha_{n-1} & b-\lambda c\\
            \alpha_{1}^2  & \cdots & \alpha_{n-1}^2 & b^2-\lambda c^2\\
            \vdots & \cdots & \vdots & \vdots \\
            \alpha_{1}^{h-1}  & \cdots & \alpha_{n-1}^{h-1} & b^{h-1}-\lambda c^{h-1}\\
            \alpha_{1}^h+\eta \alpha_{1}^{k-1+t}  & \cdots & \alpha_{n-1}^h+\eta \alpha_{n-1}^{k-1+t} & b^h-\lambda c^h+\eta(b^{k-1+t}-\lambda c^{k-1+t})\\
            \alpha_{1}^{h+1} & \cdots & \alpha_{n-1}^{h+1} & b^{h+1}-\lambda c^{h+1}\\
            \vdots & \cdots & \vdots & \vdots \\
            \alpha_{1}^{k-1}  & \cdots & \alpha_{n-1}^{k-1} & b
            ^{k-1}-\lambda c^{k-1}
        \end{bmatrix},
    \end{equation}}}
    and a generator matrix of $\textnormal{RCTRS}_{h, t}(\bm{\alpha}, b, c, \lambda, \eta, \infty)$ is given by:
    {\small{
    \begin{equation}\label{generator matrix RCTRS_inf}
        G_{h, t}(\bm \alpha,b,c,\lambda, \eta, \infty)=\begin{bmatrix}
            1 & \cdots & 1 & 1-\lambda & 0\\
            \alpha_{1} & \cdots & \alpha_{n-1} & b-\lambda c & 0\\
            \alpha_{1}^2 & \cdots & \alpha_{n-1}^2 & b^2-\lambda c^2 & 0\\
            \vdots & \cdots & \vdots & \vdots & \vdots\\
            \alpha_{1}^{h-1}  & \cdots & \alpha_{n-1}^{h-1} & b^{h-1}-\lambda c^{h-1} & 0\\
            \alpha_{1}^h+\eta \alpha_{1}^{k-1+t} &  \cdots & \alpha_{n-1}^h+\eta \alpha_{n-1}^{k-1+t} & b^h-\lambda c^h+\eta(b^{k-1+t}-\lambda c^{k-1+t}) & 0\\
            \alpha_{1}^{h+1} & \cdots & \alpha_{n-1}^{h+1} & b^{h+1}-\lambda c^{h+1} & 0\\
            \vdots & \cdots & \vdots & \vdots & \vdots \\
            \alpha_{1}^{k-1}  & \cdots & \alpha_{n-1}^{k-1} & b
            ^{k-1}-\lambda c^{k-1} & 1
        \end{bmatrix}.
    \end{equation}}}
    This completes the proof.\hfill$\square$
\end{proof}
\begin{remark}\label{remark: schur sq of extended RCTRS code}
    Applying Lemma \ref{lem:: puncturing} to the extended RCTRS code and the corresponding RCTRS code yields
$$\dim\bigl(\textnormal{RCTRS}_{h, t}(\bm{\alpha}, b, c, \lambda, \eta, \infty)^{\star 2}\bigr)\ \ge\ \dim\bigl(\textnormal{RCTRS}_{h, t}(\bm{\alpha}, b, c, \lambda, \eta)^{\star 2}\bigr),$$
since puncturing the extended code at the coordinate corresponding to $f_{k-1}$ recovers the original code. We will use this observation repeatedly below; in particular, it allows the inequivalence results to be stated simultaneously for both the versions at once.
\end{remark}
\section{RCTRS and their Extended Codes at $(h,t)=(0,1)$}\label{Section 4}

In this section, we explicitly construct non-RS MDS codes $\textnormal{RCTRS}_{0, 1}(\bm{\alpha}, b, c, \lambda, \eta)$ and\\ $\textnormal{RCTRS}_{0, 1}(\bm{\alpha}, b, c, \lambda, \eta, \infty)$, where the parameters are as per Definition \ref{defn:: RCTRS}. Moreover, we prove that the codes $\textnormal{RCTRS}_{0, 1}(\bm{\alpha}, b, c, \lambda, \eta)$ and $\textnormal{RCTRS}_{0, 1}(\bm{\alpha}, b, c, \lambda, \eta, \infty)$ are not equivalent to CTRS, extended CTRS codes, or $\textnormal{TRS}_{k, 0, 1}$ codes.

Throughout the paper, we assume $[n]=\{1,2,\dots,n\}$. The following lemma gives necessary and sufficient conditions for RCTRS codes and their extended codes to be MDS.

\begin{lemma}\label{lem:: MDS condition for 0 and 1}
    \begin{enumerate}
        \item [(a)] The code $\textnormal{RCTRS}_{0, 1}(\bm{\alpha}, b, c, \lambda, \eta)$ is an MDS code if and only if the following two conditions hold:
            \begin{enumerate}
                \item [(i)] for any $\mathcal{I}\subseteq[n-1]$ of  size $k$, $(-1)^k\eta\underset{i\in \mathcal{I}}{\prod}\alpha_i\ne 1,$ and
                \item [(ii)] for any subset $\mathcal{J}\subseteq[n-1]$ of size $k-1,$ $\Phi_{\mathcal{J}}(b)\ne\lambda\Phi_{\mathcal{J}}(c),$ where $$\Phi_{\mathcal{J}}(x)=\left(1+(-1)^{|\mathcal{J}|}\eta x\prod_{j\in\mathcal{J}}\alpha_{j}\right)\prod_{j\in \mathcal{J}}(x-\alpha_{j}).$$
            \end{enumerate}
        \item [(b)] The code $\textnormal{RCTRS}_{0, 1}(\bm{\alpha}, b, c, \lambda, \eta, \infty)$ is an MDS code if and only if the following four conditions hold:
            \begin{enumerate}
                \item [(i)] for any $\mathcal{I}\subseteq[n-1]$ of  size $k$, $(-1)^k\eta\underset{i\in \mathcal{I}}{\prod}\alpha_i\ne 1,$
                \item [(ii)] for any subset $\mathcal{J}\subseteq[n-1]$ of size $k-1,$ $(-1)^{k-1}\eta\prod_{j\in \mathcal{J}}\alpha_{j}\sum_{j\in \mathcal{J}}\alpha_{j}\ne 1,$ 
                \item [(iii)] for any subset $\mathcal{J}\subseteq[n-1]$ of size $k-1,$ $\Phi_{\mathcal{J}}(b)\ne\lambda\Phi_{\mathcal{J}}(c),$ and
                \item [(iv)] for any subset $\mathcal{L}\subseteq[n-1]$ of size $k-2,$ $\Xi_{\mathcal{L}}(b)\ne \lambda\,\Xi_{\mathcal{L}}(c),$ where
                $$\Xi_{\mathcal{L}}(x)=\left(1+(-1)^{|\mathcal{L}|}\eta\, x\prod_{\ell\in\mathcal{L}}\alpha_{\ell}\Bigl(x+\sum_{\ell\in\mathcal{L}}\alpha_{\ell}\Bigr)\right)\prod_{\ell\in \mathcal{L}}(x-\alpha_{\ell}).$$
            \end{enumerate}
    \end{enumerate}
\end{lemma}
\begin{proof}
    An $[n,k]$-linear code is MDS if and only if any set of $k$ columns of its generator matrix is linearly independent. Thus, to prove part $(a)$, it is enough to show that any $k\times k$ submatrix of $G_{0,1}(\bm \alpha,b, c, \lambda, \eta)$ is invertible. We show that for any subsets $\mathcal{I}:=\{i_1, \dots, i_k\}$ and $\mathcal{J}:=\{j_1, \dots, j_{k-1}\}$ of $[n-1]$, the matrices
    $$
    A=\begin{bmatrix}
        1+\eta \alpha_{i_1}^k & 1+\eta \alpha_{i_2}^k &\cdots & 1+\eta \alpha_{i_{k}}^k\\
            \alpha_{i_1} & \alpha_{i_2} & \cdots & \alpha_{i_{k}}\\
            \alpha_{i_1}^2 & \alpha_{i_2}^2 & \cdots & \alpha_{i_{k}}^2\\
            \vdots& \vdots & \cdots & \vdots \\
            \alpha_{i_1}^{k-1} & \alpha_{i_2}^{k-1} & \cdots & \alpha_{i_{k}}^{k-1}
    \end{bmatrix}
    $$
    and 
    $$
    B=\begin{bmatrix}
        1+\eta \alpha_{j_1}^k & 1+\eta \alpha_{j_2}^k &\cdots & 1+\eta \alpha_{j_{k-1}}^k & 1-\lambda +\eta(b^k-\lambda c^k)\\
            \alpha_{j_1} & \alpha_{j_2} & \cdots & \alpha_{j_{k-1}} & b-\lambda c\\
            \alpha_{j_1}^2 & \alpha_{j_2}^2 & \cdots & \alpha_{j_{k-1}}^2 & b^2-\lambda c^2\\
            \vdots& \vdots & \cdots & \vdots & \vdots \\
            \alpha_{j_1}^{k-1} & \alpha_{j_2}^{k-1} & \cdots & \alpha_{j_{k-1}}^{k-1} & b^{k-1}-\lambda c^{k-1}
    \end{bmatrix}
    $$
    are both invertible. Following the proof of Lemma 4 of \cite{beelen2017twisted}, it is clear that $\det A\ne 0$ if and only if $(-1)^k\eta\underset{i\in \mathcal{I}}{\prod}\alpha_i\ne 1$. In order to compute $\det B$, it is clear that
    {\small{
    \begin{align*}
        \det B&=\begin{vmatrix}
        1 & 1 &\cdots & 1 & 1-\lambda \\
            \alpha_{j_1} & \alpha_{j_2} & \cdots & \alpha_{j_{k-1}} & b-\lambda c\\
            \alpha_{j_1}^2 & \alpha_{j_2}^2 & \cdots & \alpha_{j_{k-1}}^2 & b^2-\lambda c^2\\
            \vdots& \vdots & \cdots & \vdots & \vdots \\
            \alpha_{j_1}^{k-1} & \alpha_{j_2}^{k-1} & \cdots & \alpha_{j_{k-1}}^{k-1} & b^{k-1}-\lambda c^{k-1}
    \end{vmatrix}+
    \eta \begin{vmatrix}
        \alpha_{j_1}^k & \alpha_{j_2}^k &\cdots & \alpha_{j_{k-1}}^k & b^k-\lambda c^k\\
            \alpha_{j_1} & \alpha_{j_2} & \cdots & \alpha_{j_{k-1}} & b-\lambda c\\
            \alpha_{j_1}^2 & \alpha_{j_2}^2 & \cdots & \alpha_{j_{k-1}}^2 & b^2-\lambda c^2\\
            \vdots& \vdots & \cdots & \vdots & \vdots \\
            \alpha_{j_1}^{k-1} & \alpha_{j_2}^{k-1} & \cdots & \alpha_{j_{k-1}}^{k-1} & b^{k-1}-\lambda c^{k-1}
    \end{vmatrix}\\ \\
    &=\Delta_1(b)-\lambda \Delta_1(c)+\eta (\Delta_2(b)-\lambda \Delta_2 (c)), \text{ where}
    \end{align*}}}
    $$\Delta_1(x):=\begin{vmatrix}
        1 & 1 &\cdots & 1 & 1 \\
            \alpha_{j_1} & \alpha_{j_2} & \cdots & \alpha_{j_{k-1}} & x\\
            \alpha_{j_1}^2 & \alpha_{j_2}^2 & \cdots & \alpha_{j_{k-1}}^2 & x^2\\
            \vdots& \vdots & \cdots & \vdots & \vdots \\
            \alpha_{j_1}^{k-1} & \alpha_{j_2}^{k-1} & \cdots & \alpha_{j_{k-1}}^{k-1} & x^{k-1}
    \end{vmatrix} \text{ and }\Delta_2(x)=\begin{vmatrix}
        \alpha_{j_1}^k & \alpha_{j_2}^k &\cdots & \alpha_{j_{k-1}}^k & x^k\\
            \alpha_{j_1} & \alpha_{j_2} & \cdots & \alpha_{j_{k-1}} & x\\
            \alpha_{j_1}^2 & \alpha_{j_2}^2 & \cdots & \alpha_{j_{k-1}}^2 & x^2\\
            \vdots& \vdots & \cdots & \vdots & \vdots \\
            \alpha_{j_1}^{k-1} & \alpha_{j_2}^{k-1} & \cdots & \alpha_{j_{k-1}}^{k-1} & x^{k-1}
    \end{vmatrix}.$$ Now, 
     $\Delta_2(x)=(-1)^{k-1}x\prod_{i=1}^{k-1}\alpha_{j_i}\Delta_1(x)$ and $\Delta_1(x)=\underset{1\le i < \ell\le k-1}{\prod}(\alpha_{j_{\ell}}-\alpha_{j_i})\prod_{i=1}^{k-1}(x-\alpha_{j_i}).$ As a result,
     {\small{
     \begin{align*}
         &\det B=\left(1+(-1)^{k-1}\eta b\prod_{i=1}^{k-1}\alpha_{j_i}\right)\Delta_1(b)-\lambda\left(1+(-1)^{k-1}\eta c\prod_{i=1}^{k-1}\alpha_{j_i}\right)\Delta_1(c)\\
         &=\underset{1\le i < \ell\le k-1}{\prod}(\alpha_{j_{\ell}}-\alpha_{j_i})\left(\left(1+(-1)^{k-1}\eta b\prod_{i=1}^{k-1}\alpha_{j_i}\right)\prod_{i=1}^{k-1}(b-\alpha_{j_i})-\lambda\left(1+(-1)^{k-1}\eta c\prod_{i=1}^{k-1}\alpha_{j_i}\right)\prod_{i=1}^{k-1}(c-\alpha_{j_i})\right)\\
         &=\underset{1\le i < \ell\le k-1}{\prod}(\alpha_{j_{\ell}}-\alpha_{j_i}) (\Phi_{\mathcal{J}}(b)-\lambda\Phi_{\mathcal{J}}(c)),
     \end{align*}}}
where $\mathcal{J}=\{j_1, j_2, \dots, j_{k-1}\}$ and 
$$\Phi_{\mathcal{J}}(x)=\left(1+(-1)^{k-1}\eta x\prod_{j\in\mathcal{J}}\alpha_{j}\right)\prod_{j\in \mathcal{J}}(x-\alpha_{j}).
$$
Thus, $\det B\ne 0$ if and only if $\Phi_{\mathcal{J}}(b)\ne \lambda\Phi_{\mathcal{J}}(c).$ Thus, we have part $(a)$.
The proof of part $(b)$ is similar to part $(a)$. We first derive condition $(ii)$. For $\textnormal{RCTRS}_{0, 1}(\bm{\alpha}, b, c, \lambda,$
$ \eta, \infty)$ to be MDS, one of the conditions is that the matrix

    $$
    D=\begin{bmatrix}
        1+\eta \alpha_{j_1}^k & 1+\eta \alpha_{j_2}^k &\cdots & 1+\eta \alpha_{j_{k-1}}^k & 0\\
            \alpha_{j_1} & \alpha_{j_2} & \cdots & \alpha_{j_{k-1}} & 0\\
            \alpha_{j_1}^2 & \alpha_{j_2}^2 & \cdots & \alpha_{j_{k-1}}^2 & 0\\
            \vdots& \vdots & \cdots & \vdots & \vdots \\
            \alpha_{j_1}^{k-1} & \alpha_{j_2}^{k-1} & \cdots & \alpha_{j_{k-1}}^{k-1} & 1
    \end{bmatrix}.
    $$ must be invertible. Using Proposition \ref{prop:: Det of vandermonde with a row deleted}, it is easy to prove that 
    \begin{align*}
        \det D&=\begin{vmatrix}
        1 & 1 &\cdots & 1 \\
            \alpha_{j_1} & \alpha_{j_2} & \cdots & \alpha_{j_{k-1}}\\
            \alpha_{j_1}^2 & \alpha_{j_2}^2 & \cdots & \alpha_{j_{k-1}}^2 \\
            \vdots& \vdots & \cdots & \vdots \\
            \alpha_{j_1}^{k-2} & \alpha_{j_2}^{k-2} & \cdots & \alpha_{j_{k-1}}^{k-2}
    \end{vmatrix}+
    (-1)^{k-2}\eta\prod_{i=1}^{k-1}\alpha_{j_i} \begin{vmatrix}
        1 & 1 &\cdots & 1 \\
            \alpha_{j_1} & \alpha_{j_2} & \cdots & \alpha_{j_{k-1}}\\
            \alpha_{j_1}^2 & \alpha_{j_2}^2 & \cdots & \alpha_{j_{k-1}}^2 \\
            \vdots& \vdots & \cdots & \vdots \\
            \alpha_{j_1}^{k-3} & \alpha_{j_2}^{k-3} & \cdots & \alpha_{j_{k-1}}^{k-3}\\
            \alpha_{j_1}^{k-1} & \alpha_{j_2}^{k-1} & \cdots & \alpha_{j_{k-1}}^{k-1}
    \end{vmatrix}\\
    &= \underset{1\le i < \ell\le k-1}{\prod}(\alpha_{j_{\ell}}-\alpha_{j_{i}})+(-1)^{k-2}\eta\prod_{i=1}^{k-1}\alpha_{j_i}\underset{1\le i < \ell\le k-1}{\prod}(\alpha_{j_{\ell}}-\alpha_{j_{i}})\sum_{i=1}^{k-1} \alpha_{j_i}.
\end{align*}

    Consequently, $\det D\ne 0$ if and only if $(-1)^{k-2}\eta\prod_{j\in \mathcal{J}}\alpha_{j}\sum_{j\in \mathcal{J}}\alpha_{j}\ne -1,$ or equivalently, $(-1)^{k-1}\eta\prod_{j\in \mathcal{J}}\alpha_{j}\sum_{j\in \mathcal{J}}\alpha_{j}\ne 1,$  where $\mathcal{J}\subseteq [n-1]$ with size $k-1$.
    Additionally, if we consider the $k$ columns having indices in $[n-1]$, this case is equivalent to conditions $(i)$ and $(iii)$.

    Finally, suppose that the chosen $k$ columns consist of the last two columns of $G_{0,1}(\bm\alpha, b, c, \lambda, \eta,$ $ \infty)$ together with the $k-2$ columns indexed by $\mathcal{L}=\{\ell_1, \dots, \ell_{k-2}\}\subseteq[n-1].$ Expanding along the last column, which is $\bm e_k^{T},$ leaves the $(k-1)\times(k-1)$ determinant formed by the rows of exponents $0, 1, \dots, k-2,$ with the row of exponent $0$ replaced by $1+\eta x^{k},$ and by the columns indexed by $\mathcal{L}$ together with the twisted column. Splitting that row by linearity and splitting the twisted column as $b^{\bullet}-\lambda c^{\bullet}$ exactly as in part $(a)$, we are led to the determinant
    $$\begin{vmatrix}
        \alpha_{\ell_1}^{k} & \cdots & \alpha_{\ell_{k-2}}^{k} & x^{k}\\
        \alpha_{\ell_1} & \cdots & \alpha_{\ell_{k-2}} & x\\
        \vdots & \cdots & \vdots & \vdots\\
        \alpha_{\ell_1}^{k-2} & \cdots & \alpha_{\ell_{k-2}}^{k-2} & x^{k-2}
    \end{vmatrix}
    =(-1)^{k-2}\, x\prod_{\ell\in\mathcal{L}}\alpha_{\ell}\Bigl(x+\sum_{\ell\in\mathcal{L}}\alpha_{\ell}\Bigr)\det V(\alpha_{\ell_1}, \dots, \alpha_{\ell_{k-2}}, x),$$
    To see this, move the first row to the last position, which contributes the sign $(-1)^{k-2},$ and then take the factor $\alpha_{\ell_1}, \dots, \alpha_{\ell_{k-2}}, x$ out of the respective columns; what remains carries the exponents $0, 1, \dots, k-3, k-1$ on the $k-1$ points $\alpha_{\ell_1}, \dots, \alpha_{\ell_{k-2}}, x,$ and Proposition \ref{prop:: Det of vandermonde with a row deleted} applied with $n=k-1$ and $h=k-2,$ evaluates it as $\sigma_1(\alpha_{\ell_1}, \dots, \alpha_{\ell_{k-2}}, x)\det V(\alpha_{\ell_1}, \dots, \alpha_{\ell_{k-2}}, x).$ Consequently that determinant equals
    $$\underset{1\le i<j\le k-2}{\prod}(\alpha_{\ell_j}-\alpha_{\ell_i})\bigl(\Xi_{\mathcal{L}}(b)-\lambda\,\Xi_{\mathcal{L}}(c)\bigr),$$
    and it is non-zero precisely when condition $(iv)$ holds. \hfill$\square$
\end{proof}
A construction of non-RS MDS codes based on RCTRS and its extended code is shown in the following theorem.
\begin{theorem}\label{thm:: MDS code 0 and 1 construction subfield}
    Let $\mathbb{F}_{q_0}\subsetneq \mathbb{F}_{q_1}\subset\mathbb{F}_{q}$ be a chain of subfields of $\mathbb{F}_{q}.$ Suppose that $b, c\in\mathbb{F}_{q_0}$ and for $i\in[n-1],~\alpha_i$ be distinct in $\mathbb{F}_{q_0}$, where $n\le q_0$.
    If $\lambda\in \mathbb{F}_{q_1}\setminus\mathbb{F}_{q_0}$ and $\eta\in\mathbb{F}_{q}\setminus\mathbb{F}_{q_1}^{*},$ then the codes $\textnormal{RCTRS}_{0, 1}(\bm{\alpha}, b, c, \lambda, \eta)$ and $\textnormal{RCTRS}_{0, 1}(\bm{\alpha}, b, c, \lambda, \eta, \infty)$ are both MDS. 
    Moreover, if $b,c\in\mathbb{F}_{q_0}^*$ with $b\ne c$, and either $\eta=0$ or $\eta^2 \notin \mathbb{F}_{q_0},$ with $4\le k\le \frac{n}{2}$, then the codes $\textnormal{RCTRS}_{0, 1}(\bm{\alpha}, b, c, \lambda, \eta)$ and $\textnormal{RCTRS}_{0, 1}(\bm{\alpha}, b, c, \lambda, \eta, \infty)$ are both non-RS MDS codes.
\end{theorem}
\begin{proof}
    We first prove that the code $\textnormal{RCTRS}_{0, 1}(\bm{\alpha}, b, c, \lambda, \eta)$ is MDS. 
    Suppose that $\mathcal{I}\subseteq [n-1]$ such that $|\mathcal{I}|=k$ and $(-1)^k\eta\underset{i\in \mathcal{I}}{\prod}\alpha_i= 1$. 
    Then, $\underset{i\in\mathcal{I}}{\prod}\alpha_i\in\mathbb{F}_{q_0}^{*}$, and hence $\eta=(-1)^k \underset{i\in\mathcal{I}}{\prod}\alpha_i^{-1}\in\mathbb{F}_{q_0}^{*}$, a contradiction.
    Thus, for every $\mathcal{I}\subseteq [n-1]$ such that $|\mathcal{I}|=k$, we have $(-1)^k\eta\underset{i\in \mathcal{I}}{\prod}\alpha_i\ne 1$. 
    Next, assume that $\mathcal{J}=\{i_1,i_2,\dots,i_{k-1}\}\subseteq [n-1]$ such that $\Phi_{\mathcal{J}}(b)=\lambda\Phi_{\mathcal{J}}(c)$, where $\Phi$ is defined in Lemma \ref{lem:: MDS condition for 0 and 1}. Then,
    \begin{equation*}
        \left(1+(-1)^{k-1}\eta b\prod_{j=1}^{k-1}\alpha_{i_j}\right)\prod_{j=1}^{k-1}(b-\alpha_{i_j})=\lambda\left(1+(-1)^{k-1}\eta c\prod_{j=1}^{k-1}\alpha_{i_j}\right)\prod_{j=1}^{k-1}(c-\alpha_{i_j}).
    \end{equation*}
    If $\alpha_{j}=0,$ for some $j\in\mathcal{J},$ then $\lambda=\prod_{j\in\mathcal{J}}\left(\frac{b-\alpha_{j}}{c-\alpha_{j}}\right)\in \mathbb{F}_{q_0},$ a contradiction. Hence, $\alpha_{j}\ne 0,$ for all $j\in\mathcal{J},$ and if $z=(-1)^{k-1}\eta  \prod_{j\in \mathcal{J}}\alpha_{j},$ then
    \begin{equation}\label{eqn:: expression for z 0 1}
        z=\frac{\lambda\prod_{j\in\mathcal{J}}(c-\alpha_{j})-\prod_{j\in\mathcal{J}}(b-\alpha_{j})}{b\prod_{j\in\mathcal{J}}(b-\alpha_{j})-\lambda c\prod_{j\in\mathcal{J}}(c-\alpha_{j})}
     \end{equation}
     Since $\alpha_i, b, c,\in \mathbb{F}_{q_0}$ and $\lambda\in \mathbb{F}_{q_1}$, we must have $z\in \mathbb{F}_{q_1}$, provided the denominator of the RHS of \eqref{eqn:: expression for z 0 1} is non-zero, which is already true as if the denominator vanishes, 
     we have $\lambda=\prod_{j\in\mathcal{J}}\left(\frac{b-\alpha_{j}}{c-\alpha_{j}}\right)\in \mathbb{F}_{q_0},$ a contradiction. As a result, $z\in\mathbb{F}_{q_1}$, and in fact, $z\in\mathbb{F}_{q_1}^{*}.$ Hence, $\eta\in\mathbb{F}_{q_1}^*,$ again a contradiction. Therefore, for every subset $\mathcal{J}\subseteq [n-1]$ of size $k-1$, $\Phi_{\mathcal{J}}(b)\ne\lambda\Phi_{\mathcal{J}}(c)$. The result now follows from Lemma \ref{lem:: MDS condition for 0 and 1}.
     
    Now, we determine $\dim \mathcal{C}^{\star 2}$, for $b\neq c$, where $\mathcal{C}:=\textnormal{RCTRS}_{0, 1}(\bm{\alpha}, b, c, \lambda, \eta).$ If $\bm{g}_1, \bm{g}_2, \dots, \bm{g}_k$ denote the $k$ rows of the generator matrix (\ref{generator matrix RCTRS}) of $\mathcal{C},$ then
    $$\mathcal{D}:=\{\bm{g}_1\star\bm{g}_1,\, \bm{g}_1\star\bm{g}_2\} \cup\{\bm{g}_2\star\bm{g}_2,\, \bm{g}_2\star\bm{g}_3,\, \dots ,\bm{g}_2\star\bm{g}_k\}\cup \{\bm{g}_3\star\bm{g}_k,\, \dots,\, \bm{g}_k\star\bm{g}_k\}\cup\{\bm{g}_3\star\bm{g}_3\}$$
    is a subset of $\mathcal{C}^{\star 2}$ containing $2k$ distinct vectors. We show that $\mathcal{D}$ is linearly independent. Consider
     $$
     D=\begin{bmatrix}
         1+\eta^2\alpha_1^{2k}+2\eta\alpha_1^{k} & \cdots & 1+\eta^2\alpha_{n-1}^{2k}+2\eta\alpha_{n-1}^{k} & (1-\lambda+\eta(b^k-\lambda c^k))^2\\
         \alpha_1+\eta\alpha_1^{k+1} & \cdots& \alpha_{n-1}+\eta\alpha_{n-1}^{k+1} & (1-\lambda+\eta(b^k-\lambda c^k))(b-\lambda c)\\
         \alpha_1^2 & \cdots& \alpha_{n-1}^2 & (b-\lambda c)(b-\lambda c)\\
         \vdots & \cdots & \vdots & \vdots\\
         \alpha_1^{2k-2} & \cdots& \alpha_{n-1}^{2k-2} & (b^{k-1}-\lambda c^{k-1})(b^{k-1}-\lambda c^{k-1})\\
         \alpha_1^4 & \cdots& \alpha_{n-1}^4 & (b^2-\lambda c^2)(b^2-\lambda c^2)\\
     \end{bmatrix},
     $$
     whose rows are the vectors in $\mathcal{D}$, listed in the order in which they appear in the definition of $\mathcal{D}$. Observe that $D$ is row-equivalent to
     $$
     \begin{bmatrix}
         1+\eta^2\alpha_1^{2k} & \cdots & 1+\eta^2\alpha_{n-1}^{2k} & *\\
         \alpha_1 & \cdots& \alpha_{n-1} & *\\
         \alpha_1^2 & \cdots& \alpha_{n-1}^2 & (b-\lambda c)(b-\lambda c)\\
         \vdots & \cdots & \vdots & \vdots\\
         \alpha_1^{2k-2} & \cdots& \alpha_{n-1}^{2k-2} & (b^{k-1}-\lambda c^{k-1})(b^{k-1}-\lambda c^{k-1})\\
         0 & \cdots& 0 & \lambda bc (b-c)^{2}
     \end{bmatrix},
     $$
     where the last row operation is $R_{2k}\rightarrow R_{2k}-R_5$.
    Its $2k\times 2k$ submatrix
     $$
     D'=\begin{bmatrix}
         1+\eta^2\alpha_1^{2k} & \cdots & 1+\eta^2\alpha_{2k-1}^{2k} & *\\
         \alpha_1 & \cdots& \alpha_{2k-1} & *\\
         \alpha_1^2 & \cdots& \alpha_{2k-1}^2 & *\\
         \vdots & \cdots & \vdots & \vdots\\
         \alpha_1^{2k-2} & \cdots& \alpha_{2k-1}^{2k-2} & *\\
         0 & \cdots& 0 & \lambda bc (b-c)^{2}
     \end{bmatrix}
     $$
     has rank $2k$ if and only if $\lambda b c (b-c)^2(1+\eta^2 \prod_{j=1}^{2k-1} \alpha_j \sum_{j=1}^{2k-1} \alpha_j)\ne 0.$ 
     Since $\alpha_i\in\mathbb{F}_{q_0}$ and either $\eta=0$ or $\eta^2\notin\mathbb{F}_{q_0},$ $1+\eta^2 \prod_{j=1}^{2k-1} \alpha_j \sum_{j=1}^{2k-1} \alpha_j$ is always non-zero. 
     Since $\lambda, b,c$ are all non-zero with $b\neq c$, rank of $D$ is  $2k$.  
     Consequently, $\dim (\mathcal{C}^{\star 2})\ge 2k.$ Hence $\mathcal{C}$ is a non-RS MDS code by Remark \ref{remark:: non RS MDS code}.
     That $\textnormal{RCTRS}_{0, 1}(\bm{\alpha}, b, c, \lambda, \eta, \infty)$ is MDS can be proved in a similar way. The non-RS property of $\textnormal{RCTRS}_{0, 1}(\bm{\alpha}, b, c, \lambda, \eta, \infty)$ follows from Remarks \ref{remark:: non RS MDS code} and \ref{remark: schur sq of extended RCTRS code}. \hfill$\square$ 
\end{proof}
\begin{example}
    Consider the chain of subfields $\mathbb{F}_{11}\subsetneq \mathbb{F}_{11^2}\subsetneq\mathbb{F}_{11^4},$ and let $\gamma\in \mathbb{F}_{11^2}$ and $\zeta \in \mathbb{F}_{11^4}$ both be primitive. For each $i\in [7],$ let $\alpha_i=i-1,$ so that $n=8\le 11.$ Take $b=8,$ $c=7,$ $\lambda=\gamma,$ $\eta=\zeta$ and $k=4.$ Here $\lambda\in\mathbb{F}_{11^2}\setminus\mathbb{F}_{11}$ and $\eta\in\mathbb{F}_{11^4}\setminus\mathbb{F}_{11^2}^{*},$ and $\eta^2\notin\mathbb{F}_{11}$. Also $b, c\in\mathbb{F}_{11}^{*}$ with $b\ne c,$ and $4\le k\le \frac{n}{2}=4.$ Theorem \ref{thm:: MDS code 0 and 1 construction subfield} therefore applies, and the codes $\textnormal{RCTRS}_{0, 1}(\bm{\alpha}, b, c, \lambda, \eta)$ and $\textnormal{RCTRS}_{0, 1}(\bm{\alpha}, b, c, \lambda, \eta, \infty)$ are non-RS MDS codes with parameters $[8,4,5]$ and $[9,4,6]$ over $\mathbb{F}_{11^4},$ respectively. A computation in \textsc{Magma} \cite{Magma} confirms that both codes are MDS and that their Schur squares have dimensions $8$ and $9,$ respectively, each of which differs from $2k-1=7.$
\end{example}
  The next theorem gives another explicit construction of non-RS MDS codes $\textnormal{RCTRS}_{0, 1}(\bm{\alpha}, b,$
  $ c,\lambda, \eta)$ and $\textnormal{RCTRS}_{0, 1}(\bm{\alpha}, b, c, \lambda, \eta, \infty)$. By virtue of this construction, we can construct RCTRS codes with larger lengths as compared to Theorem \ref{thm:: MDS code 0 and 1 construction subfield}.
 \begin{theorem}\label{thm:: MDS code 0 and 1 construction subgroup}
     Let $\mathbb{F}_{q_0}$ be a subfield of $\mathbb{F}_q$, and let $G$ be a subgroup of $\mathbb{F}_{q_0}^{*}$ of order $n$. For $b, c\in\mathbb{F}_{q_0}$ with $b\ne c$ and $G=\{1, \mu_1, \mu_2, \dots, \mu_{n-1}\},$ define $\alpha_{i}:=\frac{b-\mu_i c}{1-\mu_i},$ for $i\in [n-1]$. 
     If $\lambda\in \mathbb{F}_{q_0}\setminus G$ and $\eta\in \mathbb{F}_{q}\setminus\mathbb{F}_{q_0}^{*},$ then the codes $\textnormal{RCTRS}_{0, 1}(\bm{\alpha}, b, c, \lambda, \eta)$ and $\textnormal{RCTRS}_{0, 1}(\bm{\alpha}, b, c, $
     $\lambda, \eta, \infty)$ are both MDS.
     Moreover, if $b,c, \lambda$ are all non-zero, and either $\eta=0$ or $\eta^2 \notin \mathbb{F}_{q_0},$ with $4 \le k\le \frac{n}{2},$ then the codes $\textnormal{RCTRS}_{0, 1}(\bm{\alpha}, b, c, \lambda, \eta)$ and $\textnormal{RCTRS}_{0, 1}(\bm{\alpha}, b, c, \lambda, \eta, \infty)$ are both non-RS MDS code.
 \end{theorem}
 \begin{proof}
     We prove the result for $\textnormal{RCTRS}_{0, 1}(\bm{\alpha}, b, c, \lambda, \eta)$ and a proof for $\textnormal{RCTRS}_{0, 1}(\bm{\alpha}, b, c, \lambda,$
     $ \eta, \infty)$ follows similarly.

     Note that the evaluation points $\alpha_i$ are pairwise distinct. 
     For, if $\alpha_i=\alpha_j$ for some $i\ne j \in [n-1],$ then $b=c,$ a contradiction. 
     Observe that for any $i\in[n-1],$ $\alpha_i\ne b.$ Suppose that there is a subset $\mathcal{I}\subseteq [n-1]$ of size $k$ such that $(-1)^k\eta\underset{i\in \mathcal{I}}{\prod}\alpha_i= 1$. 
     Hence, $\alpha_i\in \mathbb{F}_{q_0}^{*}$ for all $i\in\mathcal{I}$ and consequently, $\underset{i\in\mathcal{I}}{\prod}\alpha_i\in\mathbb{F}_{q_0}^{*}$. As a result, $\eta=(-1)^k \underset{i\in\mathcal{I}}{\prod}\alpha_i^{-1}\in\mathbb{F}_{q_0}^{*}$, a contradiction. Therefore, for every subset $\mathcal{I}\subseteq [n-1]$ of size $k$, we must have $(-1)^k\eta\underset{i\in \mathcal{I}}{\prod}\alpha_i\ne 1$. Next, suppose that there is a subset $\mathcal{J}\subseteq [n-1]$ of size $k-1$ such that $\Phi_{\mathcal{J}}(b)=\lambda\Phi_{\mathcal{J}}(c)$ holds. Then,
    \begin{equation*}
        \left(1+(-1)^{k-1}\eta b\prod_{j=1}^{k-1}\alpha_{i_j}\right)\prod_{j=1}^{k-1}(b-\alpha_{i_j})=\lambda\left(1+(-1)^{k-1}\eta c\prod_{j=1}^{k-1}\alpha_{i_j}\right)\prod_{j=1}^{k-1}(c-\alpha_{i_j}).
    \end{equation*}
    If $\alpha_{j}=0,$ for some $j\in\mathcal{J},$ then $\lambda=\prod_{j\in\mathcal{J}}\left(\frac{b-\alpha_{j}}{c-\alpha_{j}}\right)=\prod_{j\in\mathcal{J}} \mu_{j}\in G,$ a contradiction. If $\alpha_{j}\ne 0,$ for any $j\in\mathcal{J},$ and if $z=(-1)^{k-1}\eta  \prod_{j\in \mathcal{J}}\alpha_{j},$ then
    \begin{equation}\label{eqn:: expression for z 0 1 subgroup}
        z=\frac{\lambda\prod_{j\in\mathcal{J}}(c-\alpha_{j})-\prod_{j\in\mathcal{J}}(b-\alpha_{j})}{b\prod_{j\in\mathcal{J}}(b-\alpha_{j})-\lambda c\prod_{j\in\mathcal{J}}(c-\alpha_{j})}
     \end{equation}
     Note that the denominator of the RHS of Equation \eqref{eqn:: expression for z 0 1 subgroup} is non-zero. If it was zero, then $\lambda=\prod_{j\in\mathcal{J}}\left(\frac{b-\alpha_{j}}{c-\alpha_{j}}\right)=\prod_{j\in\mathcal{J}}\mu_{j}\in G,$ a contradiction. Since $\alpha_i, b, c,\lambda\in \mathbb{F}_{q_0}$, we must have $z\in \mathbb{F}_{q_0}$. In fact, $z\in\mathbb{F}_{q_0}^{*}$. 
     As a result, $\eta\in\mathbb{F}_{q_0}^{*}$, again a contradiction. Therefore, for every subset $\mathcal{J}\subseteq [n-1]$ of size $k-1$, $\Phi_{\mathcal{J}}(b)\ne\lambda\Phi_{\mathcal{J}}(c)$. The result now follows from Lemma \ref{lem:: MDS condition for 0 and 1}.

    A proof that they are non-RS MDS codes is similar to the proof in Theorem \ref{thm:: MDS code 0 and 1 construction subfield}. \hfill$\square$   
 \end{proof}
 \begin{remark}
     If $q$ is odd, Theorem \ref{thm:: MDS code 0 and 1 construction subfield} produces non-RS MDS codes of length at most $q$ (or $q+1$) over $\mathbb{F}_{q^4}.$ However, Theorem \ref{thm:: MDS code 0 and 1 construction subgroup} produces non-RS MDS codes of length $\frac{q-1}{2}$ (or $\frac{q+1}{2}$) over $\mathbb{F}_{q^2}$ and hence, non-RS MDS codes of length $\frac{q^2-1}{2}$ (or $\frac{q^2+1}{2}$) over $\mathbb{F}_{q^4}.$
 \end{remark}
 \begin{corollary}
     There exists non-RS MDS row-column twisted RS codes of length $\frac{q-1}{p}$ and $\frac{q-1}{p}+1$ over $\mathbb{F}_{q^2},$ where $p$ is a prime divisor of $q-1$. In particular, if $q$ is odd, then there exists non-RS MDS  row-column twisted RS codes of length $\frac{q-1}{2}$ and $\frac{q+1}{2}$ over $\mathbb{F}_{q^2}.$
 \end{corollary}
 \begin{proof}
     There exists a subgroup of order $\frac{q-1}{p}$ of $\mathbb{F}_q^{*},$ for every prime divisor $p$ of $q-1$. The result now follows from Theorem \ref{thm:: MDS code 0 and 1 construction subgroup}.\hfill$\square$
 \end{proof}
 \begin{remark}
     Theorem \ref{thm:: MDS code 0 and 1 construction subgroup} also allows for $\eta$ to be $0.$ In such a case, the codes $\textnormal{RCTRS}_{0, 1}(\bm{\alpha}, b, c,$
     $ \lambda, \eta)$ and $\textnormal{RCTRS}_{0, 1}(\bm{\alpha}, b, c, \lambda, \eta, \infty)$ can be considered over $\mathbb{F}_{q_0}$ instead of $\mathbb{F}_q.$ In fact, this is Theorem 1 of \cite{liu2025column}. 
 \end{remark}
 \begin{example}
     Consider the field $\mathbb{F}_{23^2}$ with $\gamma$ as its primitive element. Then $\mathbb{F}_{23}$ is a proper subfield of $\mathbb{F}_{23^2}$ and $G:=\langle 2 \rangle=\{1, 2, 3, 4, 6, 8, 9, 12, 13, 16, 18\}$ is a subgroup of $\mathbb{F}_{23}^{*}$ of order $11.$ Let $b=12$ and $c=7,$ then by Theorem \ref{thm:: MDS code 0 and 1 construction subgroup}, the evaluation points $\alpha_i$, for $i\in [10],$ are $2, 3, 4, 6, 13, 15, 16, 17, 20, 22$. Let $\lambda= 5, \eta=\gamma$ and $k=4.$ Then the codes $\textnormal{RCTRS}_{0, 1}(\bm{\alpha}, b, c, \lambda, \eta)$ and $\textnormal{RCTRS}_{0, 1}(\bm{\alpha}, b, c, \lambda, \eta, \infty)$ have parameters $[11, 4, 8]$ and $[12, 4, 9]$, and are both non-RS MDS codes over $\mathbb{F}_{23^2}$ by Theorem \ref{thm:: MDS code 0 and 1 construction subgroup}. The computations are verified using \textsc{Magma} \cite{Magma}.
 \end{example}
 The next example shows that one may obtain non-RS MDS codes for certain choices of $\eta\in\mathbb{F}_{q_0}^{*}$, hence, obtaining non-RS MDS row-column twisted codes of larger lengths.
 \begin{example}
     Consider the finite field $\mathbb{F}_{17}$ and $G:=\langle 2\rangle=\{1, 2, 4, 8, 9, 13, 15, 16\}$ be a subgroup of $\mathbb{F}_{17}^{*}$ of order $8.$ Let $b=1$ and $c=2.$ Suppose that the evaluation points $\alpha_i$, for $i\in [7],$ are $0, 3, 7, 8, 10, 12, 13,$ which are calculated as in Theorem \ref{thm:: MDS code 0 and 1 construction subgroup}. Let $\lambda= 10, \eta=4$ and $k=4.$ Then by \textsc{Magma} \cite{Magma}, the code $\textnormal{RCTRS}_{0, 1}(\bm{\alpha}, b, c, \lambda, \eta)$  is an $[8, 4, 5]$ linear code over $\mathbb{F}_{17}.$ It can be easily checked that the dimension of the Schur square of $\textnormal{RCTRS}_{0, 1}(\bm{\alpha}, b, c, \lambda, \eta)$ is $2k,$ hence, $\textnormal{RCTRS}_{0, 1}(\bm{\alpha}, b, c, \lambda, \eta)$ is a non-RS $[8, 4]$-MDS code over $\mathbb{F}_{17}.$
 \end{example}
  By computing the Schur squares of the codes $\textnormal{RCTRS}_{0, 1}(\bm{\alpha}, b, c, \lambda, \eta)$ and $\textnormal{RCTRS}_{0, 1}(\bm{\alpha}, b, c, $ $\lambda, \eta),$ we show that, under mild assumptions, these codes are not equivalent to any column-twisted RS code, extended column-twisted RS code, or the corresponding TRS code (that is, a TRS codes with $h=0$ and $t=1$) under mild assumptions. Thus, our construction yields new families of non-RS MDS codes beyond those considered in \cite{liu2025column} and some of the families in \cite{beelen2017twisted}.
 \begin{theorem}\label{thm:: non-equivalence of MDS code 0 and 1 construction subgroup and CTRS}
     Let $\mathbb{F}_{q_0}$ be a proper subfield of $\mathbb{F}_q,$ and let $G$ be a subgroup of $\mathbb{F}_{q_0}^{*}$ of order $n$. For $b, c\in\mathbb{F}_{q_0}^{*}$ with $b\ne c$ and $G=\{1, \mu_1, \mu_2, \dots, \mu_{n-1}\},$ define $\alpha_{i}:=\frac{b-\mu_i c}{1-\mu_i},$ for $i\in [n-1]$. 
     If $\lambda\in \mathbb{F}_{q_0}^{*}\setminus G,\eta\in \mathbb{F}_{q}\setminus\mathbb{F}_{q_0}$ with $\eta^2 \notin \mathbb{F}_{q_0},$ and $4\le k\le \frac{n-1}{2}$, then the codes $\textnormal{RCTRS}_{0, 1}(\bm{\alpha}, b, c, \lambda, \eta)$ and $\textnormal{RCTRS}_{0, 1}(\bm{\alpha}, b, c, \lambda, \eta, \infty)$ are both non-RS MDS codes and are not equivalent to any CTRS or extended CTRS code. Moreover, neither code is equivalent to any TRS code with hook $h=0$ and twist $t=1.$
 \end{theorem}
 \begin{proof}
     We prove the result for $\mathcal{C}:=\textnormal{RCTRS}_{0, 1}(\bm{\alpha}, b, c, \lambda, \eta)$ and a proof for $\textnormal{RCTRS}_{0, 1}(\bm{\alpha}, b, c,$ $ \lambda, \eta, \infty)$ follows similarly. 
     By Theorem \ref{thm:: MDS code 0 and 1 construction subgroup}, the code $\mathcal{C}$ is an $[n, k]$ non-RS MDS codes over $\mathbb{F}_q.$ Its generator matrix $G:=G_{0,1}(\bm{\alpha}, b, c, \lambda, \eta)$ is given by plugging $h=0$ and $t=1$ in  \eqref{generator matrix RCTRS}. If we label the rows of the generator matrix as $\bm{g}_i,$ for all $i\in[k],$ then 
     \begin{equation*}
         \begin{split}
             \mathcal{D}:=\{\bm{g}_1\star\bm{g}_1,\, \bm{g}_1\star\bm{g}_2\}\cup  \{\bm{g}_2\star\bm{g}_2,\, \bm{g}_2\star\bm{g}_3,\, \dots ,\bm{g}_2\star\bm{g}_k\}\\\cup \{\bm{g}_3\star\bm{g}_k,\, \bm{g}_4\star\bm{g}_k \dots,\, \bm{g}_k\star\bm{g}_k\}\cup \{\bm{g}_1\star\bm{g}_k, \bm{g}_3\star\bm{g}_3\}
         \end{split}
     \end{equation*}
          contains $2k+1$ distinct vectors and is contained in $\mathcal{C}^{\star 2}$. We show that $\mathcal{D}$ is linearly independent. Consider,
     {\small{
     \begin{equation*}
         D=\begin{bmatrix}
         1+\eta^2\alpha_1^{2k}+2\eta\alpha_1^{k} & \cdots & 1+\eta^2\alpha_{n-1}^{2k}+2\eta\alpha_{n-1}^{k} & (1-\lambda+\eta(b^k-\lambda c^k))^2\\
         \alpha_1+\eta\alpha_1^{k+1} & \cdots& \alpha_{n-1}+\eta\alpha_{n-1}^{k+1} & (1-\lambda+\eta(b^k-\lambda c^k))(b-\lambda c)\\
         \alpha_1^2 & \cdots& \alpha_{n-1}^2 & (b-\lambda c)(b-\lambda c)\\
         \vdots & \cdots & \vdots & \vdots\\
         \alpha_1^{2k-2} & \cdots& \alpha_{n-1}^{2k-2} & (b^{k-1}-\lambda c^{k-1})(b^{k-1}-\lambda c^{k-1})\\
         \alpha_1^{k-1}+\eta\alpha_1^{2k-1} & \cdots& \alpha_{n-1}^{k-1}+\eta\alpha_{n-1}^{2k-1} & ((1-\lambda)+\eta(b^k-\lambda c^k))(b^{k-1}-\lambda c^{k-1})\\
         \alpha_1^4 & \cdots& \alpha_{n-1}^4 & (b^2-\lambda c^2)(b^2-\lambda c^2)\\
        \end{bmatrix}
     \end{equation*}}}
     whose rows are the vectors in $\mathcal{D}$, listed in the order in which they appear in the definition of $\mathcal{D}$. Then $D$ is row-equivalent to
     $$
     \begin{bmatrix}
         1+\eta^2\alpha_1^{2k} & \cdots & 1+\eta^2\alpha_{n-1}^{2k}& *\\
         \alpha_1 & \cdots& \alpha_{n-1} & *\\
         \alpha_1^2 & \cdots& \alpha_{n-1}^2 & *\\
         \vdots & \cdots & \vdots & \vdots\\
         \alpha_1^{2k-2} & \cdots& \alpha_{n-1}^{2k-2} & *\\
         \eta\alpha_1^{2k-1} & \cdots& \eta\alpha_{n-1}^{2k-1} & *\\
         0 & \cdots& 0 & \lambda b c(b-c)^2\\
     \end{bmatrix}.
     $$
     Consider its $(2k+1)\times (2k+1)$ submatrix
     $$
     D'=\begin{bmatrix}
         1+\eta^2\alpha_1^{2k} & \cdots & 1+\eta^2\alpha_{2k}^{2k}& *\\
         \alpha_1 & \cdots& \alpha_{2k} & *\\
         \alpha_1^2 & \cdots& \alpha_{2k}^2 & *\\
         \vdots & \cdots & \vdots & \vdots\\
         \alpha_1^{2k-2} & \cdots& \alpha_{2k}^{2k-2} & *\\
         \eta\alpha_1^{2k-1} & \cdots& \eta\alpha_{2k}^{2k-1} & *\\
         0 & \cdots& 0 & \lambda b c(b-c)^2\\
     \end{bmatrix}.
     $$
     Then $\det D'\ne 0$ if and only if $\lambda b c (b-c)^2 \eta (1-\eta^2 \prod_{i=1}^{2k}\alpha_i)\ne 0.$ Under the hypothesis, the last quantity is non-zero and hence, $\textnormal{rank}(D)= 2k+1.$ 
     Consequently, $\dim (\mathcal{C}^{\star 2})\ge 2k+1.$

     On the other hand, by \cite[Theorems 1 and 2]{liu2025column} the Schur square of a column-twisted Reed-Solomon code $\textnormal{CTRS}(\bm{\alpha}, b', c', \lambda')$, and of its extended code, has dimension exactly $2k$ whenever $b'\ne c'$ and $\lambda'\ne 0.$ The argument given there for the value $2k$ uses only that the evaluation points are pairwise distinct, and so applies to any choice of parameters. If $\lambda'=0$ or $b'=c',$ then $\textnormal{CTRS}(\bm{\alpha}', b', c', \lambda')$ is a GRS code, whose Schur square has dimension $2k-1.$ Lemma \ref{cor:: schur square bounds}(a) gives the matching upper bound $2k$ in every case. Hence the Schur square of an $[n,k]$-MDS CTRS code, or of an extended CTRS code, has dimension $2k-1$ or $2k,$ and by Lemma \ref{cor:: schur square bounds}(c) the Schur square of a $\textnormal{TRS}_{k, 0, 1}$ code has dimension at most $2k.$ Since $\dim(\mathcal{C}^{\star 2})\ge 2k+1,$ Remark \ref{remark:: non RS MDS code} shows that $\mathcal{C}$ is inequivalent to every CTRS code, every extended CTRS code, and every $\textnormal{TRS}_{k, 0, 1}$ code over $\mathbb{F}_q.$\hfill$\square$
 \end{proof}
 \begin{example}
     Consider the field $\mathbb{F}_{31^2}$ with $\gamma$ as its primitive element. Then $\mathbb{F}_{31}$ is a proper subfield of $\mathbb{F}_{31^2}$ and $G:=\langle 18 \rangle=\{1, 18, 19, 2, 20, 4, 5, 7, 25, 8, 9, 10, 28, 14, 16\}$ is a subgroup of $\mathbb{F}_{31}^{*}$ of order $15.$ Let $b=1$ and $c=5,$ then by Theorem \ref{thm:: MDS code 0 and 1 construction subgroup}, the evaluation points $\alpha_i$, for $i\in [14],$ are $18, 19, 9, 15, 27, 6, 16, 0, 10, 21, 2, 4, 22, 28$. Let $\lambda= 11, \eta=\gamma$ and $k=5.$ Then the codes $\textnormal{RCTRS}_{0, 1}(\bm{\alpha}, b, c, \lambda, \eta)$ and $\textnormal{RCTRS}_{0, 1}(\bm{\alpha}, b, c, \lambda, \eta, \infty)$ are MDS codes with parameters $[15,5,11]$ and $[16,5,12]$, respectively, over $\mathbb{F}_{31^2}$. Using \textsc{Magma} \cite{Magma}, we verify that the Schur squares of both codes have dimension $11$. Consequently, both codes are non-Reed--Solomon MDS codes and are inequivalent to CTRS, extended CTRS, and TRS codes with $h=0$ and $t=1$. This is consistent with Theorem \ref{thm:: non-equivalence of MDS code 0 and 1 construction subgroup and CTRS}.
 \end{example}
\section{RCTRS and their Extended Codes at $(h,t)=(k-1,1)$}\label{Section 5}

In this section, we explicitly construct non-RS MDS codes $\textnormal{RCTRS}_{k-1, 1}(\bm{\alpha}, b, c, \lambda, \eta)$ and\\ $\textnormal{RCTRS}_{k-1, 1}(\bm{\alpha}, b, c, \lambda, \eta, \infty)$ that are not equivalent to any CTRS or extended CTRS code.
\begin{lemma}\label{lem:: MDS condition k-1 1} 
    \begin{enumerate}
        \item [(a)] The code $\textnormal{RCTRS}_{k-1, 1}(\bm{\alpha}, b, c, \lambda, \eta)$ is an MDS code if and only if the following two conditions hold:
            \begin{enumerate}
                \item [(i)] for any $\mathcal{I}\subseteq[n-1]$ of  size $k$, $\eta\sum_{i\in\mathcal{I}} \alpha_i\ne -1$, and
                \item [(ii)] for any subset $\mathcal{J}\subseteq[n-1]$ of size $k-1,$ $\Psi_{\mathcal{J}}(b)\ne\lambda\Psi_{\mathcal{J}}(c),$ where $$\Psi_{\mathcal{J}}(x):=\left(1+\eta x+ \eta\sum_{j\in\mathcal{J}}\alpha_j\right)\prod_{j\in\mathcal{J}}(x-\alpha_{j}).$$
            \end{enumerate}
        \item [(b)] The code $\textnormal{RCTRS}_{k-1, 1}(\bm{\alpha}, b, c, \lambda, \eta, \infty)$ is an MDS code if and only if the following three conditions hold:
            \begin{enumerate}
                \item [(i)] for any $\mathcal{I}\subseteq[n-1]$ of  size $k$, $\eta\sum_{i\in\mathcal{I}} \alpha_i\ne -1,$
                \item [(ii)] for any subset $\mathcal{J}\subseteq[n-1]$ of size $k-1,$ $\Psi_{\mathcal{J}}(b)\ne\lambda\Psi_{\mathcal{J}}(c),$ and 
                \item [(iii)] for any subset $\mathcal{L}\subseteq[n-1]$ of size $k-2,\, \Omega_{\mathcal{L}}(b)\ne \lambda\Omega_{\mathcal{L}}(c),$ where
                $$
                    \Omega_{\mathcal{L}}(x)=\prod_{\ell\in\mathcal{L}}(x-\alpha_{\ell}).
                $$
            \end{enumerate}
    \end{enumerate}
\end{lemma}
\begin{proof}
    (a). It suffices to show that for any subset $\mathcal{I}:=\{i_1, \dots, i_k\}$ and $\mathcal{J}:=\{j_1, \dots, j_{k-1}\}$ of $[n-1]$, the matrices
    $$
    A=\begin{bmatrix}
            1 & 1 &\cdots & 1 \\
            \alpha_{i_1} & \alpha_{i_2} & \cdots & \alpha_{i_{k}}\\
            \alpha_{i_1}^2 & \alpha_{i_2}^2 & \cdots & \alpha_{i_{k}}^2 \\
            \vdots& \vdots & \cdots & \vdots \\
            \alpha_{i_1}^{k-2} & \alpha_{i_2}^{k-2} & \cdots & \alpha_{i_{k}}^{k-2}\\
            \alpha_{i_1}^{k-1}+\eta \alpha_{i_1}^{k} & \alpha_{i_2}^{k-1}+\eta \alpha_{i_2}^{k} & \cdots & \alpha_{i_{k}}^{k-1}+\eta \alpha_{i_{k}}^{k} \\
        \end{bmatrix} \text{ and}
    $$
    
    $$
    B=\begin{bmatrix}
            1 & 1 &\cdots & 1 & 1-\lambda\\
            \alpha_{j_1} & \alpha_{j_2} & \cdots & \alpha_{j_{k-1}} & b-\lambda c\\
            \alpha_{j_1}^2 & \alpha_{j_2}^2 & \cdots & \alpha_{j_{k-1}}^2 & b^2-\lambda c^2\\
            \vdots& \vdots & \cdots & \vdots & \vdots \\
            \alpha_{j_1}^{k-2} & \alpha_{j_2}^{k-2} & \cdots & \alpha_{j_{k-1}}^{k-2} & b^{k-2}-\lambda c^{k-2}\\
            \alpha_{j_1}^{k-1}+\eta \alpha_{j_1}^{k} & \alpha_{j_2}^{k-1}+\eta \alpha_{j_2}^{k} & \cdots & \alpha_{j_{k-1}}^{k-1}+\eta \alpha_{j_{k-1}}^{k} & b^{k-1}-\lambda c^{k-1} +\eta(b^k-\lambda c^k)\\
        \end{bmatrix}
    $$
    are both invertible. It is easy to prove that $\det A\ne 0$ if and only if $\eta\sum_{i\in\mathcal{I}}\alpha_{i}\ne -1$. In order to compute $\det B$, it is clear that $\det B=\Delta_1(b)-\lambda\Delta_1(c)+\eta (\Delta_2(b)-\lambda\Delta_2(c))$, where 
    \begin{align*}
        \Delta_1(x)=\begin{vmatrix}
            1 & 1 &\cdots & 1 & 1\\
            \alpha_{j_1} & \alpha_{j_2} & \cdots & \alpha_{j_{k-1}} & x\\
            \alpha_{j_1}^2 & \alpha_{j_2}^2 & \cdots & \alpha_{j_{k-1}}^2 & x^2\\
            \vdots& \vdots & \cdots & \vdots & \vdots \\
            \alpha_{j_1}^{k-1} & \alpha_{j_2}^{k-1} & \cdots & \alpha_{j_{k-1}}^{k-1} & x^{k-1}
        \end{vmatrix}, \text{ and } 
        \Delta_2(x)=\begin{vmatrix}
            1 & 1 &\cdots & 1 & 1\\
            \alpha_{j_1} & \alpha_{j_2} & \cdots & \alpha_{j_{k-1}} & x\\
            \alpha_{j_1}^2 & \alpha_{j_2}^2 & \cdots & \alpha_{j_{k-1}}^2 & x^2\\
            \vdots& \vdots & \cdots & \vdots & \vdots \\
            \alpha_{j_1}^{k-2} & \alpha_{j_2}^{k-2} & \cdots & \alpha_{j_{k-1}}^{k-2} & x^{k-2}\\
            \alpha_{j_1}^{k} &  \alpha_{j_2}^{k} & \cdots & \alpha_{j_{k-1}}^{k} & x^{k}\\
        \end{vmatrix}.
    \end{align*}
    Now using Proposition \ref{prop:: Det of vandermonde with a row deleted}, we have
    \begin{align*}
        \Delta_2(x)&= \sigma_{1}(\alpha_{j_1}, \alpha_{j_2}, \dots, \alpha_{j_{k-1}}, x) \det V(\alpha_{j_1}, \dots, \alpha_{j_{k-1}}, x)\\
        &=\left(x+\sum_{j\in\mathcal{J}}\alpha_j\right)\det V(\alpha_{j_1}, \dots, \alpha_{j_{k-1}}, x),
    \end{align*}
    where $V(x_1, \dots x_n)=(x_i^{j-1})_{1\le i,j\le n}$ denotes the Vandermonde matrix and $ \sigma_{r}(X_1, X_2, \dots, X_N)$ denotes the $r$-th elementary symmetric functions in $N$ variables.

    Hence,
    \begin{align*}
        \det B&= \det V(\alpha_{j_1}, \dots, \alpha_{j_{k-1}}, b)\left(1+\eta\left(b+\sum_{j\in\mathcal{J}}\alpha_j\right)\right)\\
        &-\lambda\det V(\alpha_{j_1}, \dots, \alpha_{j_{k-1}}, c)\left(1+\eta\left(c+\sum_{j\in\mathcal{J}}\alpha_j\right)\right)\\
        &=\underset{1\le i < \ell\le k-1}{\prod}(\alpha_{j_{\ell}}-\alpha_{j_i}) (\Psi_{\mathcal{J}}(b)-\lambda \Psi_{\mathcal{J}}(c)),
    \end{align*}
     where
    $\Psi_{\mathcal{J}}(x):=\prod_{i=1}^{k-1}(x-\alpha_{j_i})\left(1+\eta x+ \eta\sum_{j\in\mathcal{J}}\alpha_j\right).
    $
    Consequently, $\det B\ne 0$ if and only if $\Psi_{\mathcal{J}}(b)\ne \lambda \Psi_{\mathcal{J}}(c).$ This proves part $(a)$.
    
    Part $(b)$ can be proved in a similar manner.\hfill$\square$
 \end{proof}
A construction of non-RS MDS $\textnormal{RCTRS}_{k-1, 1}(\bm{\alpha}, b, c, \lambda, \eta)$ and $\textnormal{RCTRS}_{k-1, 1}(\bm{\alpha}, b, c, \lambda, \eta, \infty)$ is given in the next theorem.
 \begin{theorem}\label{thm:: MDS code k-1 and 1 construction subgroup}
     Let $\mathbb{F}_{q_0}$ be a subfield of $\mathbb{F}_q,$ and let $G$ be a subgroup of $\mathbb{F}_{q_0}^{*}$ of order $n$. For $b, c\in\mathbb{F}_{q_0}$ with $b\ne c$ and $G=\{1, \mu_1, \mu_2, \dots, \mu_{n-1}\},$ define $\alpha_{i}:=\frac{b-\mu_i c}{1-\mu_i},$ for $i\in [n-1]$. If $\lambda\in \mathbb{F}_{q_0}\setminus G$ and $\eta\in \mathbb{F}_{q}\setminus\mathbb{F}_{q_0}^{*},$ then the codes $\textnormal{RCTRS}_{k-1, 1}(\bm{\alpha}, b, c, \lambda, \eta)$ and $\textnormal{RCTRS}_{k-1, 1}(\bm{\alpha}, b, c, \lambda, \eta, \infty)$ are both MDS with parameters $[n, k, n-k+1]$ and $[n+1, k, n-k+2]$, respectively. Moreover, if $\lambda$ and $\eta$ are non-zero, and $4 \le k\le \frac{n}{2},$ then the codes $\textnormal{RCTRS}_{k-1, 1}(\bm{\alpha}, b, c, \lambda, \eta)$ and $\textnormal{RCTRS}_{k-1, 1}(\bm{\alpha}, b, c, \lambda, \eta, \infty)$ are both non-RS MDS codes.
 \end{theorem}
 \begin{proof}
 We prove that $\textnormal{RCTRS}_{k-1, 1}(\bm{\alpha}, b, c, \lambda, \eta, \infty)$ is MDS; the MDS property of\\ $\textnormal{RCTRS}_{k-1, 1}(\bm{\alpha}, b, c,\lambda, \eta)$ follows similarly. Suppose that there is a subset $\mathcal{I}\subseteq [n-1]$ of size $k$ such that $\eta \sum_{i\in \mathcal{I}}\alpha_i=-1.$ 
 Being non-zero, $\sum_{i\in \mathcal{I}}\alpha_i\in\mathbb{F}_{q_0}^*$, and hence $\eta=-\left(\sum_{i\in \mathcal{I}}\alpha_i\right)^{-1}\in \mathbb{F}_{q_0}^*,$ a contradiction.
 Suppose that there is subset $\mathcal{J}\subseteq [n-1]$ of size $k-1$ such that $\Psi_{\mathcal{J}}(b)=\lambda\Psi_{\mathcal{J}}(c),$ i.e., 
 $$
    \prod_{j\in\mathcal{J}}(b-\alpha_{j})\left(1+\eta b+ \eta\sum_{j\in\mathcal{J}}\alpha_j\right)=\lambda\prod_{j\in\mathcal{J}}(c-\alpha_{j})\left(1+\eta c+ \eta\sum_{j\in\mathcal{J}}\alpha_j\right).
 $$
 Then
 \begin{equation*}
     \eta=\frac{\lambda\prod_{j\in\mathcal{J}}(c-\alpha_{j})-\prod_{j\in\mathcal{J}}(b-\alpha_{j})}{(b+ \sum_{j\in\mathcal{J}}\alpha_j)\prod_{j\in\mathcal{J}}(b-\alpha_{j})-\lambda(c+ \sum_{j\in\mathcal{J}}\alpha_j)\prod_{j\in\mathcal{J}}(c-\alpha_{j})}.
 \end{equation*}
  It is easy to see that $\eta\ne 0$ and the denominator of the RHS of the above equation is non-zero, otherwise $\lambda=\prod_{j\in\mathcal{J}}\left(\frac{b-\alpha_j}{c-\alpha_j}\right)\in G,$ a contradiction. 
  Since $\alpha_i$, $b, c\in\mathbb{F}_{q_0},$ we must have $\eta\in\mathbb{F}_{q_0}^{*},$ again a contradiction to the hypothesis. Lastly, suppose that there is a subset $\mathcal{L}\subseteq[n-1]$ of size $k-2$ such that $\prod_{\ell\in\mathcal{L}}(b-\alpha_{\ell})=\lambda\prod_{\ell\in\mathcal{L}}(c-\alpha_{\ell}).$ This cannot hold since $\lambda\notin G.$ The result now follows from Lemma \ref{lem:: MDS condition k-1 1}. 
  
  A proof that the codes $\textnormal{RCTRS}_{k-1, 1}(\bm{\alpha}, b, c, \lambda)$ and $\textnormal{RCTRS}_{k-1, 1}(\bm{\alpha}, b, c, \lambda, \eta, \infty)$ are non-RS MDS codes can be established in the same manner as in Theorem \ref{thm:: MDS code 0 and 1 construction subgroup}. In fact, if the rows of the generator matrix are $\bm{g}_i,$ for all $i\in[k],$ then 
    \begin{equation*}
        \begin{split}
            \mathcal{D}:=\{\bm{g}_1\star\bm{g}_1,\, \bm{g}_1\star\bm{g}_2,\, \dots,  \bm{g}_1\star\bm{g}_{k-1}\}\cup \{\bm{g}_2\star\bm{g}_{k-1},\,
    \bm{g}_3\star\bm{g}_{k-1},\dots,
    \bm{g}_{k-1}\star\bm{g}_{k-1}\}\\
    \cup
    \{\bm{g}_{k-2}\star\bm{g}_k,\, \bm{g}_{k-1}\star\bm{g}_k, \bm{g}_2\star\bm{g}_2\}
        \end{split}
    \end{equation*}
    is a linearly independent subset of $\mathcal{C}^{\star 2}$ of size $2k$.\hfill$\square$
 \end{proof}
 We next construct a family of non-RS MDS $\textnormal{RCTRS}_{k-1, 1}(\bm{\alpha}, b, c, \lambda, \eta)$ and $\textnormal{RCTRS}_{k-1, 1}(\bm{\alpha}, b, c, $
 $\lambda, \eta, \infty)$ codes which are not equivalent to any column-twisted RS codes or extended column-twisted RS codes.
 \begin{theorem}\label{thm:: non-equivalence k-1 and 1 and CTRS}
     Suppose that $\Char(\mathbb{F}_q)$ is odd. Let $\mathbb{F}_{q_0}$ be a proper subfield of $\mathbb{F}_q,$ and let $G$ be a subgroup of $\mathbb{F}_{q_0}^{*}$ of order $n$. For $b, c\in\mathbb{F}_{q_0}$ with $b\ne c$ and $G=\{1, \mu_1, \mu_2, \dots, \mu_{n-1}\},$ define $\alpha_{i}:=\frac{b-\mu_i c}{1-\mu_i},$ for $i\in [n-1]$. If $\lambda\in \mathbb{F}_{q_0}^{*}\setminus G, \eta\in \mathbb{F}_{q}\setminus\mathbb{F}_{q_0}$ and $4\le k\le \frac{n-1}{2},$ then the codes $\textnormal{RCTRS}_{k-1, 1}(\bm{\alpha}, b, c, \lambda, \eta)$ and $\textnormal{RCTRS}_{k-1, 1}(\bm{\alpha}, b, c, \lambda, \eta, \infty)$ are both not equivalent to any CTRS or extended CTRS code.
 \end{theorem}
 \begin{proof}
    We prove the result for $\mathcal{C}:=\textnormal{RCTRS}_{k-1, 1}(\bm{\alpha}, b, c, \lambda, \eta).$  Using Theorem \ref{thm:: MDS code k-1 and 1 construction subgroup}, the code $\mathcal{C}$ is an $[n, k]$ non-RS MDS codes over $\mathbb{F}_q.$ Its generator matrix $G:=G_{k-1,1}(\bm{\alpha}, b, c, \lambda, \eta)$ is given by plugging $h=k-1$ and $t=1$ in  \eqref{generator matrix RCTRS}. If we label the rows of the generator matrix as $\bm{g}_i,$ for all $i\in[k],$ then 
    \begin{equation*}
        \begin{split}
            \mathcal{D}:=\{\bm{g}_1\star\bm{g}_1,\, \bm{g}_1\star\bm{g}_2,\, \dots,  \bm{g}_1\star\bm{g}_{k-1}\}\cup \{\bm{g}_2\star\bm{g}_{k-1},\,
    \bm{g}_3\star\bm{g}_{k-1},\dots,
    \bm{g}_{k-1}\star\bm{g}_{k-1}\}\\\cup
    \{\bm{g}_{k-2}\star\bm{g}_k,\, \bm{g}_{k-1}\star\bm{g}_k, \bm{g}_{k}\star\bm{g}_k, \bm{g}_2\star\bm{g}_2\}
        \end{split}
    \end{equation*}
    has size $2k+1$ and is contained in $\mathcal{C}^{\star 2}$. We show that $\mathcal{D}$ is linearly independent. Consider
     $$
     D=\begin{bmatrix}
         1 & \cdots & 1 & (1-\lambda)^2\\
         \alpha_1 & \cdots& \alpha_{n-1} & (1-\lambda)(b-\lambda c)\\
         \alpha_1^2 & \cdots& \alpha_{n-1}^2 & (1-\lambda)(b^2-\lambda c^2)\\
         \vdots & \cdots & \vdots & \vdots\\
         \alpha_1^{2k-4} & \cdots& \alpha_{n-1}^{2k-4} & *\\
         \alpha_1^{2k-4}+\eta\alpha_1^{2k-3} & \cdots& \alpha_{n-1}^{2k-4}+\eta\alpha_{n-1}^{2k-3} & *\\
         \alpha_1^{2k-3}+\eta\alpha_1^{2k-2} & \cdots& \alpha_{n-1}^{2k-3}+\eta\alpha_{n-1}^{2k-2} & *\\
         \alpha_1^{2k-2}+\eta^2\alpha_1^{2k}+2\eta\alpha_1^{2k-1} & \cdots& \alpha_{n-1}^{2k-1}+\eta^2\alpha_{n-1}^{2k}+2\eta\alpha_{n-1}^{2k-1} & *\\
         \alpha_1^2 & \cdots& \alpha_{n-1}^2 & (b-\lambda c)^2\\
     \end{bmatrix},
     $$
     whose rows are the vectors in $\mathcal{D}$, listed in the order in which they appear in the definition of $\mathcal{D}$. Then $D$ is row-equivalent to
     $$
     \begin{bmatrix}
         1 & \cdots & 1 & (1-\lambda)^2\\
         \alpha_1 & \cdots& \alpha_{n-1} & (1-\lambda)(b-\lambda c)\\
         \vdots & \cdots & \vdots & \vdots\\
         \alpha_1^{2k-4} & \cdots& \alpha_{n-1}^{2k-4} & *\\
         \eta\alpha_1^{2k-3} & \cdots& \eta\alpha_{n-1}^{2k-3} & *\\
         \eta\alpha_1^{2k-2} & \cdots& \eta\alpha_{n-1}^{2k-2} & *\\
         \eta^2\alpha_1^{2k}+2\eta\alpha_1^{2k-1} & \cdots& \eta^2\alpha_{n-1}^{2k}+2\eta\alpha_{n-1}^{2k-1} & *\\
         0 & \cdots& 0 & \lambda(b-c)^2\\
     \end{bmatrix}.
     $$
     Consider its $(2k+1)\times (2k+1)$ submatrix
     $$
     D'=\begin{bmatrix}
         1 & \cdots & 1& *\\
         \alpha_1 & \cdots& \alpha_{2k} & *\\
         \alpha_1^2 & \cdots& \alpha_{2k}^2 & *\\
         \vdots & \cdots & \vdots & \vdots\\
         \alpha_1^{2k-4} & \cdots& \alpha_{2k}^{2k-4} & *\\
         \eta\alpha_1^{2k-3} & \cdots& \eta\alpha_{2k}^{2k-3} & *\\
         \eta\alpha_1^{2k-2} & \cdots& \eta\alpha_{2k}^{2k-2} & *\\
         \eta^2\alpha_1^{2k}+2\eta\alpha_1^{2k-1} & \cdots& \eta^2\alpha_{2k}^{2k}+2\eta\alpha_{2k}^{2k-1} & *\\
         0 & \cdots& 0 & \lambda (b-c)^2\\
     \end{bmatrix}.
     $$
     Then $\det D'\ne 0$ if and only if $\lambda(b-c)^2\eta^3\left(\eta\sum_{i=1}^{2k}\alpha_i+2\right)\ne 0.$ Under the hypothesis, the last quantity is non-zero and hence, $\textnormal{rank}(D)\ge 2k+1.$ Consequently, $\dim (\mathcal{C}^{\star 2})\ge 2k+1.$ On the other hand, the Schur square of an $[n,k]$-MDS CTRS code, and that of an extended CTRS code, has dimension $2k-1$ or $2k,$ as recorded in the proof of Theorem \ref{thm:: non-equivalence of MDS code 0 and 1 construction subgroup and CTRS}, and by Lemma \ref{cor:: schur square bounds}(c) the Schur square of a $\textnormal{TRS}_{k, 0, 1}$ code has dimension at most $2k$. The result now follows from Remark \ref{remark:: non RS MDS code}.\hfill$\square$
 \end{proof}

 \begin{example}
    Consider the finite field $\mathbb{F}_{29}$ and $G:=\{1, 20, 4, 5, 22, 6, 23, 24, 7, 25, 9, 28, $
    $13, 16\}$ be a subgroup of $\mathbb{F}_{29}^{*}$ of order $14.$ Let $b=12$ and $c=7.$ Suppose that the evaluation points $\alpha_i$, for $i\in [13],$ are $22, 15, 13, 4, 6, 16, 3, 11, 8, 10, 24, 9, 26$ which are calculated as in Theorem \ref{thm:: MDS code k-1 and 1 construction subgroup}. Let $\lambda= 15$, let $\eta$ be a primitive element of $\mathbb{F}_{29^2}$, and let $k=4.$ A computation in \textsc{Magma} shows that the code $\textnormal{RCTRS}_{k-1, 1}(\bm{\alpha}, b, c, \lambda, \eta)$ is an $[14,4,11]$ MDS code over $\mathbb{F}_{29^2}$, while the code $\textnormal{RCTRS}_{k-1, 1}(\bm{\alpha}, b, c, \lambda, \eta, \infty)$ is an $[15,4,12]$ MDS code over $\mathbb{F}_{29^2}$. Moreover, \textsc{Magma} \cite{Magma} verifies that the Schur square of each of these codes has dimension $9$. Hence, both codes are non-Reed--Solomon MDS codes and are inequivalent to any CTRS or extended CTRS code over $\mathbb{F}_{29^2}$.
 \end{example}
\section{RCTRS and their Extended Codes for a General Hook at $t=1$}\label{Section 6}
In this section, we explicitly construct non-RS MDS codes $\textnormal{RCTRS}_{h, 1}(\bm{\alpha}, b, c, \lambda, \eta)$, for any $1\le h\le k-2.$ We also show that under mild assumptions, for $1\le h\le k-2,$ the codes $\textnormal{RCTRS}_{h, 1}(\bm{\alpha}, b, c, \lambda, \eta)$ are not equivalent to any CTRS codes, extended CTRS codes or TRS codes with twist $t=1$.
\begin{lemma}\label{lem:: MDS condition for h and 1}
    \begin{enumerate}
        \item [(a)] The code $\textnormal{RCTRS}_{h, 1}(\bm{\alpha}, b, c, \lambda, \eta)$ is an MDS code if and only if the following two conditions hold:
            \begin{enumerate}
                \item [(i)] for any $\mathcal{I}:=\{i_1, \dots, i_k\}\subseteq[n-1]$ of  size $k$, $(-1)^{k-h}\eta\sigma_{k-h}(\alpha_{i_1}, \dots, \alpha_{i_k})\ne 1$ and
                \item [(ii)] for any subset $\mathcal{J}:=\{i_1, \dots i_{k-1}\}\subseteq[n-1]$ of size $k-1,$ $\Phi_{\mathcal{J}, h}(b)\ne\lambda\Phi_{\mathcal{J}, h}(c),$ where $$\Phi_{\mathcal{J}, h}(x):=\left(1+(-1)^{k-1-h}\eta \sigma_{k-h}(\alpha_{i_1}, \alpha_{i_2}, \dots, \alpha_{i_{k-1}}, x)\right)\prod_{j=1}^{k-1}(x-\alpha_{i_j}).$$
            \end{enumerate}
        \item [(b)] Suppose $0\le h\le k-2.$ For a subset $\mathcal{A}=\{a_1, \dots, a_s\}\subseteq[n-1]$ write
        $$\tau_{\mathcal{A}}:=\sigma_{k-1-h}(\alpha_{a_1}, \dots, \alpha_{a_s})\,\sigma_{1}(\alpha_{a_1}, \dots, \alpha_{a_s})-\sigma_{k-h}(\alpha_{a_1}, \dots, \alpha_{a_s}),$$
        and let $\tau_{\mathcal{A}}(x)$ denote the same expression evaluated at the enlarged list $\alpha_{a_1}, \dots, \alpha_{a_s}, x.$ Then $\textnormal{RCTRS}_{h, 1}(\bm{\alpha}, b, c, \lambda, \eta, \infty)$ is an MDS code if and only if conditions \textup{(i)} and \textup{(ii)} hold together with
        \begin{enumerate}
            \item [(iii)] for any subset $\mathcal{J}\subseteq[n-1]$ of size $k-1,$ $(-1)^{k-1-h}\eta\,\tau_{\mathcal{J}}\ne 1,$ and
            \item [(iv)] for any subset $\mathcal{L}\subseteq[n-1]$ of size $k-2,$ $\Xi_{\mathcal{L}, h}(b)\ne\lambda\,\Xi_{\mathcal{L}, h}(c),$ where
            $$\Xi_{\mathcal{L}, h}(x):=\bigl(1+(-1)^{k-2-h}\eta\,\tau_{\mathcal{L}}(x)\bigr)\prod_{\ell\in\mathcal{L}}(x-\alpha_{\ell}).$$
        \end{enumerate}
        The remaining case $h=k-1$ is Lemma \textup{\ref{lem:: MDS condition k-1 1}(b)}.
    \end{enumerate}
\end{lemma}
\begin{proof}
    (a). It suffices to show that for any subset $\mathcal{I}:=\{i_1, \dots, i_k\}$ and $\mathcal{J}:=\{j_1, \dots, j_{k-1}\},$ the matrices
    $$
    A=\begin{bmatrix}
            1 & 1 &\cdots & 1 \\
            \alpha_{i_1} & \alpha_{i_2} & \cdots & \alpha_{i_{k}}\\
            \alpha_{i_1}^2 & \alpha_{i_2}^2 & \cdots & \alpha_{i_{k}}^2 \\
            \vdots& \vdots & \cdots & \vdots \\
            \alpha_{i_1}^h+\eta \alpha_{i_1}^{k} & \alpha_{i_2}^h+\eta \alpha_{i_2}^{k} & \cdots & \alpha_{i_{k}}^h+\eta \alpha_{i_{k}}^{k} \\
            \vdots& \vdots & \cdots & \vdots \\
            \alpha_{i_1}^{k-1} & \alpha_{i_2}^{k-1} & \cdots & \alpha_{i_{k}}^{k-1}
        \end{bmatrix} \text{ and}
    $$
    $$
    B=\begin{bmatrix}
            1 & 1 &\cdots & 1 & 1-\lambda\\
            \alpha_{j_1} & \alpha_{j_2} & \cdots & \alpha_{j_{k-1}} & b-\lambda c\\
            \alpha_{j_1}^2 & \alpha_{j_2}^2 & \cdots & \alpha_{j_{k-1}}^2 & b^2-\lambda c^2\\
            \vdots& \vdots & \cdots & \vdots & \vdots \\
            \alpha_{j_1}^h+\eta \alpha_{j_1}^{k} & \alpha_{j_2}^h+\eta \alpha_{j_2}^{k} & \cdots & \alpha_{j_{k-1}}^h+\eta \alpha_{j_{k-1}}^{k} & b^h-\lambda c^h +\eta(b^k-\lambda c^k)\\
            \vdots& \vdots & \cdots & \vdots & \vdots \\
            \alpha_{j_1}^{k-1} & \alpha_{j_2}^{k-1} & \cdots & \alpha_{j_{k-1}}^{k-1} & b^{k-1}-\lambda c^{k-1}
        \end{bmatrix}
    $$
    are both invertible. It is easy to see that $\det A\ne 0$ if and only if $$(-1)^{k-h}\eta\sigma_{k-h}(\alpha_{i_1}, \dots, \alpha_{i_k})\ne 1.$$ To compute $\det B$, we note that $\det B=\Delta_1(b)-\lambda\Delta_1(c)+\eta (\Delta_2(b)-\lambda\Delta_2(c))$, where 
    
    \begin{align*}
        \Delta_1(x)=\begin{vmatrix}
            1 & 1 &\cdots & 1 & 1\\
            \alpha_{j_1} & \alpha_{j_2} & \cdots & \alpha_{j_{k-1}} & x\\
            \alpha_{j_1}^2 & \alpha_{j_2}^2 & \cdots & \alpha_{j_{k-1}}^2 & x^2\\
            \vdots& \vdots & \cdots & \vdots & \vdots \\
            \alpha_{j_1}^{k-1} & \alpha_{j_2}^{k-1} & \cdots & \alpha_{j_{k-1}}^{k-1} & x^{k-1}
        \end{vmatrix}, \text{ and }
        \Delta_2(x)=\begin{vmatrix}
            1 & 1 &\cdots & 1 & 1\\
            \alpha_{j_1} & \alpha_{j_2} & \cdots & \alpha_{j_{k-1}} & x\\
            \alpha_{j_1}^2 & \alpha_{j_2}^2 & \cdots & \alpha_{j_{k-1}}^2 & x^2\\
            \vdots& \vdots & \cdots & \vdots & \vdots \\
            \alpha_{j_1}^{h-1} &  \alpha_{j_2}^{h-1} & \cdots & \alpha_{j_{k-1}}^{h-1} & x^{h-1}\\
            \alpha_{j_1}^{k} &  \alpha_{j_2}^{k} & \cdots & \alpha_{j_{k-1}}^{k} & x^{k}\\
            \alpha_{j_1}^{h+1} &  \alpha_{j_2}^{h+1} & \cdots & \alpha_{j_{k-1}}^{h+1} & x^{h+1}\\
            \vdots& \vdots & \cdots & \vdots & \vdots \\
            \alpha_{j_1}^{k-1} & \alpha_{j_2}^{k-1} & \cdots & \alpha_{j_{k-1}}^{k-1} & x^{k-1}
        \end{vmatrix}.
    \end{align*}
    Now, on interchanging rows,
    {\small{
    \begin{align*}
        \Delta_2(x)&=(-1)^{k-1-h}\begin{vmatrix}
            1 & 1 &\cdots & 1 & 1\\
            \alpha_{j_1} & \alpha_{j_2} & \cdots & \alpha_{j_{k-1}} & x\\
            \alpha_{j_1}^2 & \alpha_{j_2}^2 & \cdots & \alpha_{j_{k-1}}^2 & x^2\\
            \vdots& \vdots & \cdots & \vdots & \vdots \\
            \alpha_{j_1}^{h-1} &  \alpha_{j_2}^{h-1} & \cdots & \alpha_{j_{k-1}}^{h-1} & x^{h-1}\\
            \alpha_{j_1}^{h+1} &  \alpha_{j_2}^{h+1} & \cdots & \alpha_{j_{k-1}}^{h+1} & x^{h+1}\\
            \vdots& \vdots & \cdots & \vdots & \vdots \\
            \alpha_{j_1}^{k} & \alpha_{j_2}^{k} & \cdots & \alpha_{j_{k-1}}^{k} & x^{k}
        \end{vmatrix}\\
        &= (-1)^{k-1-h}\sigma_{k-h}(\alpha_{j_1}, \alpha_{j_2}, \dots, \alpha_{j_{k-1}}, x) \det V(\alpha_{j_1}, \dots, \alpha_{j_{k-1}}, x),
    \end{align*}
    }}
    where the last equality follows from Proposition \ref{prop:: Det of vandermonde with a row deleted}. Hence,
    \begin{align*}
        \det B&= \det V(\alpha_{j_1}, \dots, \alpha_{j_{k-1}}, b)\left(1+(-1)^{k-1-h}\eta \sigma_{k-h}(\alpha_{j_1}, \alpha_{j_2}, \dots, \alpha_{j_{k-1}}, b)\right)\\
        &-\lambda\det V(\alpha_{j_1}, \dots, \alpha_{j_{k-1}}, c)\left(1+(-1)^{k-1-h}\eta \sigma_{k-h}(\alpha_{j_1}, \alpha_{j_2}, \dots, \alpha_{j_{k-1}}, c)\right)\\
        &=\underset{1\le i < \ell\le k-1}{\prod}(\alpha_{j_{\ell}}-\alpha_{j_i}) (\Phi_{\mathcal{J}, h}(b)-\lambda \Phi_{\mathcal{J}, h}(c)), \text{ where}
    \end{align*}
    $$
    \Phi_{\mathcal{J}, h}(x):=\left(1+(-1)^{k-1-h}\eta \sigma_{k-h}(\alpha_{j_1}, \alpha_{j_2}, \dots, \alpha_{j_{k-1}}, x)\right)\prod_{i=1}^{k-1}(x-\alpha_{j_i}).
    $$
    Consequently, $\det B\ne 0$ if and only if $\Phi_{\mathcal{J}, h}(b)\ne\lambda \Phi_{\mathcal{J}, h}(c).$ This proves part $(a)$.

    (b). We first record the identity on which the argument rests. Let $p_1, \dots, p_{k-1}$ be pairwise distinct elements of $\mathbb{F}_q$ and put $e_j:=\sigma_j(p_1, \dots, p_{k-1}).$ Each $p_i$ is a root of $\prod_{\ell=1}^{k-1}(y-p_{\ell}),$ so that
    $$p_i^{\,k-1}=e_1p_i^{\,k-2}-e_2p_i^{\,k-3}+\cdots+(-1)^{k-2}e_{k-1}\qquad (1\le i\le k-1).$$
    Multiplying by $p_i$ and using the preceding relation once more to eliminate $p_i^{\,k-1}$, we obtain $p_i^{\,k}$ through $1, p_i, \dots, p_i^{\,k-2},$ and collecting terms gives
    $$p_i^{\,k}=\sum_{j=0}^{k-2}(-1)^{k-2-j}\bigl(e_1e_{k-1-j}-e_{k-j}\bigr)p_i^{\,j}\qquad (1\le i\le k-1).$$
    Let $M$ be the $(k-1)\times(k-1)$ matrix whose rows carry the exponents $0, 1, \dots, k-2$ at the points $p_1, \dots, p_{k-1},$ with the row of exponent $h$ replaced by the row of exponent $k.$ The above identity expresses the replaced row as a linear combination of the rows corresponding to the exponents $0, 1, \dots, k-2.$ Since all rows except the one corresponding to exponent $h$ are already rows of $M$, subtracting the corresponding multiples of these rows from the replaced row leaves exactly $(-1)^{k-2-h}(e_1e_{k-1-h}-e_{k-h})$ times the row corresponding to the exponent $h$. Therefore
    \begin{equation}\label{eqn:: two rows}
        \det M=(-1)^{k-2-h}\bigl(e_1e_{k-1-h}-e_{k-h}\bigr)\det V(p_1, \dots, p_{k-1}).
    \end{equation}

    We may now read off the two extra conditions. A choice of $k$ columns of $G_{h, 1}(\bm\alpha, b, c, \lambda, \eta, \infty)$ which avoids the last column reproduces conditions \textup{(i)} and \textup{(ii)} exactly as in part $(a)$, so let the last column be chosen. Being $\bm e_k^{T},$ it may be expanded, and what remains is the $(k-1)\times(k-1)$ determinant formed by the rows of exponents $0, 1, \dots, k-2,$ in which the row of exponent $h$ still carries the twist $x^{h}+\eta x^{k}.$ This is the point at which the hypothesis $h\le k-2$ is used.

    Suppose first that the remaining $k-1$ chosen columns are the evaluation columns indexed by $\mathcal{J}.$ Splitting the twisted row by linearity and applying \eqref{eqn:: two rows} to the points $\alpha_{j_1}, \dots, \alpha_{j_{k-1}}$ gives
    $$\det V(\alpha_{j_1}, \dots, \alpha_{j_{k-1}})\bigl(1+(-1)^{k-2-h}\eta\,\tau_{\mathcal{J}}\bigr),$$
    which is non-zero precisely when $(-1)^{k-1-h}\eta\,\tau_{\mathcal{J}}\ne 1.$ This is condition \textup{(iii)}.

    Suppose instead that the remaining columns are the twisted column together with the $k-2$ evaluation columns indexed by $\mathcal{L}.$ Splitting the twisted row by linearity and the twisted column as $b^{\bullet}-\lambda c^{\bullet},$ and applying \eqref{eqn:: two rows} to the points $\alpha_{\ell_1}, \dots, \alpha_{\ell_{k-2}}, x,$ gives
    $$\underset{1\le i<j\le k-2}{\prod}(\alpha_{\ell_j}-\alpha_{\ell_i})\bigl(\Xi_{\mathcal{L}, h}(b)-\lambda\,\Xi_{\mathcal{L}, h}(c)\bigr),$$
    which is non-zero precisely when condition \textup{(iv)} holds. This proves part $(b)$.\hfill$\square$
 \end{proof}
 The next theorem gives explicit constructions of  MDS codes $\textnormal{RCTRS}_{h, 1}(\bm{\alpha}, b, c,$
 $ \lambda, \eta)$ for any $1\le h\le k-2.$
 \begin{theorem}\label{thm:: MDS RCTRS code h and 1 construction subgroup}
     Let $\mathbb{F}_{q_0}$ be a subfield of $\mathbb{F}_q,$ and let $G$ be a subgroup of $\mathbb{F}_{q_0}^{*}$ of order $n$. For $b, c\in\mathbb{F}_{q_0}$ with $b\ne c$ and $G=\{1, \mu_1, \mu_2, \dots, \mu_{n-1}\},$ define $\alpha_{i}:=\frac{b-\mu_i c}{1-\mu_i},$ for $i\in [n-1]$. If $\lambda\in \mathbb{F}_{q_0}\setminus G$ and $\eta\in \mathbb{F}_{q}\setminus\mathbb{F}_{q_0}^{*},$ then $\textnormal{RCTRS}_{h, 1}(\bm{\alpha}, b, c, \lambda, \eta)$ is an $[n, k, n-k+1]$ MDS code. If in addition $h\le k-2,$ then $\textnormal{RCTRS}_{h, 1}(\bm{\alpha}, b, c, \lambda, \eta, \infty)$ is an $[n+1, k, n-k+2]$ MDS code.
 \end{theorem}
 \begin{proof}
     Note that the evaluation points $\alpha_i$ are pairwise distinct and $b\neq \alpha_i$ for all $i\in[n-1]$. 
     Suppose that there is a subset $\mathcal{I}:=\{i_1, i_2, \dots, i_k\}\subseteq [n-1]$ such that $$(-1)^{k-h}\eta\sigma_{k-h}(\alpha_{i_1}, \dots, \alpha_{i_k})
     =1.$$
     Then clearly, $\sigma_{k-h}(\alpha_{i_1}, \dots, \alpha_{i_k})\ne 0$. 
     Since $\alpha_i\in\mathbb{F}_{q_0},\; \sigma_{k-h}(\alpha_{i_1}, \dots, \alpha_{i_k})\in\mathbb{F}_{q_0}^*.$  Hence, $\eta^{-1}=(-1)^{k-h}\sigma_{k-h}(\alpha_{i_1}, \dots, \alpha_{i_k})\in \mathbb{F}_{q_0}^*,$ a contradiction. Therefore, for every subset $\mathcal{I}:=\{i_1, \dots, i_{k}\}\subseteq [n-1],$ we have $(-1)^{k-h}\eta\sigma_{k-h}(\alpha_{i_1}, \dots, \alpha_{i_k})$ $\ne 1$. 
     Next, suppose that there is a subset $\mathcal{J}:=\{j_1, j_2, \dots, j_{k-1}\}\subseteq [n-1]$ such that $\Phi_{\mathcal{J}, h}(b)=\lambda\Phi_{\mathcal{J}, h}(c)$ holds. Then,
    \begin{multline*}
        \prod_{i=1}^{k-1}(b-\alpha_{j_i})\left(1+(-1)^{k-1-h}\eta \sigma_{k-h}(\alpha_{j_1}, \alpha_{j_2}, \dots, \alpha_{j_{k-1}}, b)\right)=\\
        \lambda\prod_{i=1}^{k-1}(c-\alpha_{j_i})\left(1+(-1)^{k-1-h}\eta \sigma_{k-h}(\alpha_{j_1}, \alpha_{j_2}, \dots, \alpha_{j_{k-1}}, c)\right).
    \end{multline*}
    Therefore,
    \begin{align*}
        &(-1)^{k-h-1}\eta=\\
        &\frac{\lambda\prod_{j\in\mathcal{J}}(c-\alpha_{j})-\prod_{j\in\mathcal{J}}(b-\alpha_{j})}{\sigma_{k-h}(\alpha_{j_1}, \alpha_{j_2}, \dots, \alpha_{j_{k-1}}, b)\prod_{j\in\mathcal{J}}(b-\alpha_{j})-\lambda \sigma_{k-h}(\alpha_{j_1}, \alpha_{j_2}, \dots, \alpha_{j_{k-1}}, c)\prod_{j\in\mathcal{J}}(c-\alpha_{j})}.
     \end{align*}
     Note that the denominator of the RHS of the above equation is non-zero, as if it was zero, then {$\lambda=\prod_{j\in\mathcal{J}}\left(\frac{b-\alpha_{j}}{c-\alpha_{j}}\right)=\prod_{j\in\mathcal{J}}\mu_{j}\in G,$} a contradiction. Also note that $\eta=0$ gives same contradiction. 
     Since $\alpha_i, b, c,\lambda\in \mathbb{F}_{q_0}$, we get $\eta\in \mathbb{F}_{q_0}^{*},$ again a contradiction. 
     Therefore, for every subset $\mathcal{J}\subseteq [n-1]$ of size $k-1$, $\Phi_{\mathcal{J}, h}(b)\ne\lambda\Phi_{\mathcal{J}, h}(c)$. The result now follows from Lemma \ref{lem:: MDS condition for h and 1}$(a)$.

     Suppose now that $h\le k-2,$ and consider the two further conditions of Lemma \ref{lem:: MDS condition for h and 1}$(b)$. If $(-1)^{k-1-h}\eta\,\tau_{\mathcal{J}}=1$ for some $\mathcal{J}$ of size $k-1,$ then $\tau_{\mathcal{J}}\ne 0,$ and $\tau_{\mathcal{J}}$ lies in $\mathbb{F}_{q_0}$ because it is a polynomial with integer coefficients in the $\alpha_j,$ all of which lie in $\mathbb{F}_{q_0}.$ Hence $\eta=(-1)^{k-1-h}\tau_{\mathcal{J}}^{-1}\in\mathbb{F}_{q_0}^{*},$ contrary to the choice of $\eta.$ Next, suppose that $\Xi_{\mathcal{L}, h}(b)=\lambda\,\Xi_{\mathcal{L}, h}(c)$ for some $\mathcal{L}$ of size $k-2.$ Writing $\varepsilon:=(-1)^{k-2-h},$ $B:=\prod_{\ell\in\mathcal{L}}(b-\alpha_{\ell}),$ $C:=\prod_{\ell\in\mathcal{L}}(c-\alpha_{\ell}),$ this equation reads
     $$\varepsilon\,\eta\,\bigl(B\,\tau_{\mathcal{L}}(b)-\lambda C\,\tau_{\mathcal{L}}(c)\bigr)=\lambda C-B.$$
     Here $B, C, \tau_{\mathcal{L}}(b)$ and $\tau_{\mathcal{L}}(c)$ all lie in $\mathbb{F}_{q_0},$ and $C\ne 0$ because $c\ne\alpha_{\ell}$ for every $\ell.$ If the bracket vanishes, then $\lambda C=B,$ so that $\lambda=B/C=\prod_{\ell\in\mathcal{L}}\mu_{\ell}\in G,$ contrary to $\lambda\notin G;$ and $\eta=0$ leads to the same conclusion. Otherwise the displayed equation exhibits $\eta$ as a quotient of elements of $\mathbb{F}_{q_0},$ so $\eta\in\mathbb{F}_{q_0}^{*},$ again a contradiction. Both conditions therefore hold for every admissible $\mathcal{J}$ and $\mathcal{L},$ and Lemma \ref{lem:: MDS condition for h and 1}$(b)$ shows that the extended code is MDS. \hfill $\square$
     \end{proof}
     The next three theorems show that, under mild conditions, for $1\le h\le k-2$, the codes $\textnormal{RCTRS}_{h, 1}(\bm{\alpha}, b, c, \lambda, \eta)$ are non-RS MDS codes that are not equivalent to CTRS code, extended CTRS code or TRS code with twist $t=1$, thereby demonstrating that RCTRS codes yield genuinely new families of MDS codes.
\begin{theorem}\label{thm: non-RS TRS for any h}
     Let $\mathbb{F}_{q_0}$ be a subfield of $\mathbb{F}_q,$ and let $G$ be a subgroup of $\mathbb{F}_{q_0}^{*}$ of order $n$. For $b, c\in\mathbb{F}_{q_0}$ with $b\ne c$ and $G=\{1, \mu_1, \mu_2, \dots, \mu_{n-1}\},$ define $\alpha_{i}:=\frac{b-\mu_i c}{1-\mu_i},$ for $i\in [n-1]$. Suppose that $5\le k \le \frac{n-2}{2}$ and $2\le h\le k-3$. If $\lambda\in \mathbb{F}_{q_0}^{*}\setminus G$ and $\eta\in \mathbb{F}_{q}\setminus\mathbb{F}_{q_0}$, then the code $\textnormal{RCTRS}_{h, 1}(\bm{\alpha}, b, c, \lambda, \eta)$ is a non-RS MDS code that is not equivalent to any CTRS code, extended CTRS code or TRS code with twist $t=1$. The same holds for the extended code $\textnormal{RCTRS}_{h, 1}(\bm{\alpha}, b, c, \lambda, \eta, \infty)$.
\end{theorem}
\begin{proof}
    The fact that $\mathcal{C}:=\textnormal{RCTRS}_{h, 1}(\bm{\alpha}, b, c, \lambda, \eta)$ is MDS follows from Theorem \ref{thm:: MDS RCTRS code h and 1 construction subgroup}. To prove the remaining claim, we determine the dimension of Schur square $\mathcal{C}^{\star 2}$. For $2\le h\le k-3,$ a generator matrix $G:=G_{h,1}(\bm{\alpha}, b, c, \lambda, \eta)$ of $\mathcal{C}$ is given by plugging $t=1$ in  \eqref{generator matrix RCTRS}. If we label the rows of the generator matrix as $\bm{g}_i,$ for all $i\in[k],$ then 
    \begin{equation*}
        \begin{split}
            \mathcal{D}:=\{\bm{g}_1\star\bm{g}_1,\, \bm{g}_1\star\bm{g}_2,\, \dots,  \bm{g}_1\star\bm{g}_{k}\}\cup \{\bm{g}_2\star\bm{g}_{k},\,
    \bm{g}_3\star\bm{g}_{k},\dots,
    \bm{g}_{k}\star\bm{g}_{k}\}\\ \cup
    \{\bm{g}_{h+1}\star\bm{g}_{h+1},\, \bm{g}_{h+2}\star\bm{g}_{k-1}, \bm{g}_2\star\bm{g}_2\}
        \end{split}
    \end{equation*}
    has $2k+2$ distinct vectors and is contained in $\mathcal{C}^{\star 2}$. We show that $\mathcal{D}$ is linearly independent. Consider
     $$
     D=\begin{bmatrix}
         1 & \cdots & 1 & (1-\lambda)^2\\
         \alpha_1 & \cdots& \alpha_{n-1} & (1-\lambda)(b-\lambda c)\\
         \alpha_1^2 & \cdots& \alpha_{n-1}^2 & (1-\lambda)(b^2-\lambda c^2)\\
         \vdots & \cdots & \vdots & \vdots\\
         \alpha_1^{h}+\eta\alpha_1^{k} & \cdots& \alpha_{n-1}^{h}+\eta\alpha_{n-1}^{k} & *\\
         \alpha_1^{h+1} & \cdots& \alpha_{n-1}^{h+1} & *\\
         \vdots & \cdots & \vdots & \vdots\\
         \alpha_1^{k+h-2} & \cdots& \alpha_{n-1}^{k+h-2} & *\\
         \alpha_1^{k+h-1}+\eta\alpha_1^{2k-1} & \cdots& \alpha_{n-1}^{k+h-1}+\eta\alpha_{n-1}^{2k-1} & *\\
         \alpha_1^{k+h} & \cdots& \alpha_{n-1}^{k+h} & *\\
         \vdots & \cdots & \vdots & \vdots\\
         \alpha_1^{2k-2} & \cdots& \alpha_{n-1}^{2k-2} & *\\
         \alpha_1^{2h}+\eta^2\alpha_1^{2k}+2\eta\alpha_1^{k+h} & \cdots& \alpha_{n-1}^{2h}+\eta^2\alpha_{n-1}^{2k}+2\eta\alpha_{n-1}^{k+h}& *\\
         \alpha_1^{k+h-1} & \cdots& \alpha_{n-1}^{k+h-1}& *\\
         
         \alpha_1^2 & \cdots& \alpha_{n-1}^2 & (b-\lambda c)^2\\
     \end{bmatrix},
     $$
     whose rows are vectors in $\mathcal{D}$ in the specified order. We apply the following elementary row operations in $D$: 
     $$R_{h+1}\rightarrow R_{h+1}-\eta R_{k+1},$$ 
     $$R_{k+h}\rightarrow R_{k+h}- R_{2k+1},$$
     $$R_{2k}\rightarrow R_{2k}-2\eta R_{k+h+1},$$
     $$R_{2k}\rightarrow R_{2k}-R_{2h+1},$$
     $$R_{2k+2}\rightarrow R_{2k+2}-R_{3}.$$
     
     Write $\theta$ for the entry that the last of these operations produces in the final column; its value depends on the position of the hook. If $h\ge 3,$ then $R_3=\bm g_1\star \bm g_3,$ its polynomial part is $x^2,$ and
     $$\theta=(b-\lambda c)^2-(1-\lambda)(b^2-\lambda c^2)=\lambda(b-c)^2,$$
     which is non-zero because $\lambda\ne 0$ and $b\ne c.$ If $h=2,$ then $R_3=\bm g_1\star \bm g_{h+1},$ which the first operation has already replaced by $\bm g_1\star\bm g_{h+1}-\eta\,\bm g_2\star\bm g_{k},$ and the same computation gives
     $$\theta=\lambda(b-c)\bigl((b-c)+\eta(b^{k-1}-c^{k-1})\bigr).$$
     We claim that $\theta\ne 0$. If $b^{k-1}=c^{k-1},$ then  $\theta=\lambda(b-c)^2\ne 0$. On the other hand, if $b^{k-1}\ne c^{k-1},$ then $\eta=-(b-c)(b^{k-1}-c^{k-1})^{-1}\in\mathbb{F}_{q_0},$ since $b, c\in \mathbb{F}_{q_0}$, contradicting the hypothesis $\eta\notin \mathbb{F}_{q_0}$.

     Thus, $D$ is row-equivalent to
     $$
     \begin{bmatrix}
         1 & \cdots & 1 & (1-\lambda)^2\\
         \alpha_1 & \cdots& \alpha_{n-1} & (1-\lambda)(b-\lambda c)\\
         \alpha_1^2 & \cdots& \alpha_{n-1}^2 & (1-\lambda)(b^2-\lambda c^2)\\
         \vdots & \cdots & \vdots & \vdots\\
         \alpha_1^{h} & \cdots& \alpha_{n-1}^{h}& *\\
         \alpha_1^{h+1} & \cdots& \alpha_{n-1}^{h+1} & *\\
         \vdots & \cdots & \vdots & \vdots\\
         \alpha_1^{k+h-2} & \cdots& \alpha_{n-1}^{k+h-2} & *\\
         \eta\alpha_1^{2k-1} & \cdots& \eta\alpha_{n-1}^{2k-1} & *\\
         \alpha_1^{k+h} & \cdots& \alpha_{n-1}^{k+h} & *\\
         \vdots & \cdots & \vdots & \vdots\\
         \alpha_1^{2k-2} & \cdots& \alpha_{n-1}^{2k-2} & *\\
         \eta^2\alpha_1^{2k} & \cdots& \eta^2\alpha_{n-1}^{2k}& *\\
         \alpha_1^{k+h-1} & \cdots& \alpha_{n-1}^{k+h-1}& *\\
         0 & \cdots& 0 & \theta\\
     \end{bmatrix}.
     $$
     Consider its $(2k+2)\times (2k+2)$ submatrix
     $$
     D'=\begin{bmatrix}
         1 & \cdots & 1 & (1-\lambda)^2\\
         \alpha_1 & \cdots& \alpha_{2k+1} & (1-\lambda)(b-\lambda c)\\
         \alpha_1^2 & \cdots& \alpha_{2k+1}^2 & (1-\lambda)(b^2-\lambda c^2)\\
         \vdots & \cdots & \vdots & \vdots\\
         \alpha_1^{h} & \cdots& \alpha_{2k+1}^{h}& *\\
         \alpha_1^{h+1} & \cdots& \alpha_{2k+1}^{h+1} & *\\
         \vdots & \cdots & \vdots & \vdots\\
         \alpha_1^{k+h-2} & \cdots& \alpha_{2k+1}^{k+h-2} & *\\
         \eta\alpha_1^{2k-1} & \cdots& \eta\alpha_{2k+1}^{2k-1} & *\\
         \alpha_1^{k+h} & \cdots& \alpha_{2k+1}^{k+h} & *\\
         \vdots & \cdots & \vdots & \vdots\\
         \alpha_1^{2k-2} & \cdots& \alpha_{2k+1}^{2k-2} & *\\
         \eta^2\alpha_1^{2k} & \cdots& \eta^2\alpha_{2k+1}^{2k}& *\\
         \alpha_1^{k+h-1} & \cdots& \alpha_{2k+1}^{k+h-1}& *\\
         0 & \cdots& 0 & \theta\\
     \end{bmatrix}.
     $$
     Then $\det D'\ne 0$ if and only if $\theta\,\eta^3\ne 0$$.$ Under the hypothesis, the last quantity is non-zero and hence, $\textnormal{rank}(D)\ge 2k+2.$ Consequently, $\dim (\mathcal{C}^{\star 2})\ge 2k+2.$ On the other hand, the Schur square of an RS code has dimension $2k-1$ by Lemma \ref{lem:: schur square of GRS}, that of a CTRS code or of an extended CTRS code has dimension at most $2k$ by Lemma \ref{cor:: schur square bounds}(a), and that of a TRS code with twist $t=1$ has dimension at most $2k+1$ by Lemma \ref{cor:: schur square bounds}(b). Thus, $\mathcal{C}$ cannot be equivalent to any of these codes. The required claim now follows from Remark~\ref{remark:: non RS MDS code}.

    Finally, write $\mathcal{C}^{\infty}$ for the extended code. Deleting its last coordinate returns $\mathcal{C},$ so Lemma \ref{lem:: puncturing} gives $\dim((\mathcal{C}^{\infty})^{\star 2})\ge \dim(\mathcal{C}^{\star 2})\ge 2k+2.$ The code $\mathcal{C}^{\infty}$ is MDS by Theorem \ref{thm:: MDS RCTRS code h and 1 construction subgroup}, and the four families it is compared with have Schur squares of dimension at most $2k+1,$ so the same conclusion holds for $\mathcal{C}^{\infty}$.\hfill $\square$
   \end{proof} 
\begin{theorem}\label{thm: not CTRS TRS for any h=1}
     Let $\mathbb{F}_{q_0}$ be a subfield of $\mathbb{F}_q,$ and let $G$ be a subgroup of $\mathbb{F}_{q_0}^{*}$ of order $n$. For $b, c\in\mathbb{F}_{q_0}$ with $b\ne \pm c$ and $G=\{1, \mu_1, \mu_2, \dots, \mu_{n-1}\},$ define $\alpha_{i}:=\frac{b-\mu_i c}{1-\mu_i},$ for $i\in [n-1]$. If $\lambda\in \mathbb{F}_{q_0}^{*}\setminus G$, $\eta\in \mathbb{F}_{q}\setminus\mathbb{F}_{q_0}$, and $5\le k\le \frac{n-2}{2},$ then the code $\textnormal{RCTRS}_{1, 1}(\bm{\alpha}, b, c, \lambda, \eta)$ is a non-RS MDS code that is not equivalent to any CTRS code, extended CTRS code or TRS code with twist $t=1.$ The same holds for the extended code $\textnormal{RCTRS}_{1, 1}(\bm{\alpha}, b, c, \lambda, \eta, \infty)$.
\end{theorem}
\begin{proof}
     The fact that $\mathcal{C}:=\textnormal{RCTRS}_{1,1}(\bm{\alpha},b,c,\lambda,\eta)$ is MDS follows from Theorem~\ref{thm:: MDS RCTRS code h and 1 construction subgroup}. The remaining claims follow by an argument analogous to the proof of Theorem \ref{thm: non-RS TRS for any h}. In particular, if the rows of a generator matrix of $\mathcal{C}$ are denoted by $\bm{g}_1,\ldots,\bm{g}_k$, then
    \begin{equation*}
        \begin{split}
            \mathcal{D}:=\{\bm{g}_1\star\bm{g}_1,\, \bm{g}_1\star\bm{g}_2,\, \dots,  \bm{g}_1\star\bm{g}_{k}\}\cup \{\bm{g}_2\star\bm{g}_{k},\,
    \bm{g}_3\star\bm{g}_{k},\dots,
    \bm{g}_{k}\star\bm{g}_{k}\}\\ \cup
    \{\bm{g}_{2}\star\bm{g}_{2},\, \bm{g}_{3}\star\bm{g}_{3}, \bm{g}_3\star\bm{g}_{k-1}\}
        \end{split}
    \end{equation*}
    has $2k+2$ distinct vectors and is a linearly independent subset of $\mathcal{C}^{\star 2}$. The final-coordinate contribution is obtained from $\bm g_3\star\bm g_3-\bm g_1\star \bm g_5,$ whose entry in the final coordinate is $\lambda(b^2-c^2)^2.$ This is non-zero since $\lambda\ne0$ and $b\ne\pm c$. Note that $\bm g_1\star\bm g_5$ is among the chosen Schur products because $k\ge5$. Since the Schur square of any RS code, CTRS code, extended CTRS code, or TRS code with twist $t=1$ has dimension at most $2k+1$ by Lemma \ref{lem:: schur square of GRS} and Lemma \ref{cor:: schur square bounds}, the claimed inequivalence follows from Remark \ref{remark:: non RS MDS code}. 
    
    Finally, write $\mathcal{C}^{\infty}$ for the extended code. Deleting its last coordinate returns $\mathcal{C},$ so Lemma \ref{lem:: puncturing} gives $\dim((\mathcal{C}^{\infty})^{\star 2})\ge \dim(\mathcal{C}^{\star 2})\ge 2k+2.$ The code $\mathcal{C}^{\infty}$ is MDS by Theorem \ref{thm:: MDS RCTRS code h and 1 construction subgroup}, and the families it is compared with have Schur squares of dimension at most $2k+1,$ so the same conclusion holds for $\mathcal{C}^{\infty}$.
\hfill $\square$
 \end{proof}
\begin{example}
    Consider the finite field $\mathbb{F}_{31}$ and let $G:=\langle 18 \rangle=\{1, 18, 19, 2, 20, 4, 5, 7, 25, 8, 9,$ $ 10, 28, 14, 16\}$ be a subgroup of $\mathbb{F}_{31}^{*}$ of order $15.$ Let $b=1$ and $c=5$. The evaluation points $\alpha_i, i\in[14],$ obtained as in Theorem \ref{thm:: MDS RCTRS code h and 1 construction subgroup}, are $18,19,9,15,27,6,16,0,10,21,2,4,22,28.$ Let $\lambda= 11,$ let $\eta$ be a primitive element of $\mathbb{F}_{31^2}$, and let $k=5.$ A computation in \textsc{Magma} shows that both $\textnormal{RCTRS}_{2, 1}(\bm{\alpha}, b, c, \lambda, \eta)$ and $\textnormal{RCTRS}_{1, 1}(\bm{\alpha}, b, c, \lambda, \eta)$ are $[15, 5, 11]$ MDS codes over $\mathbb{F}_{31^2}$, while their extended versions, $\textnormal{RCTRS}_{2, 1}(\bm{\alpha}, b, c, \lambda, \eta, \infty)$ and $\textnormal{RCTRS}_{1, 1}(\bm{\alpha}, b, c, \lambda, \eta, \infty)$, are $[16,5,12]$ MDS codes over $\mathbb{F}_{31^2}.$ Moreover, \textsc{Magma} \cite{Magma} verifies that the Schur square of each of the two non-extended codes has dimension $12$, whereas the Schur square of each of the two extended codes has dimension $13$. Hence, all four codes are non-Reed--Solomon MDS codes and are inequivalent to any CTRS, extended CTRS, or TRS code with $t=1$ over $\mathbb{F}_{31^2}$. This is consistent with Theorems \ref{thm: non-RS TRS for any h} and \ref{thm: not CTRS TRS for any h=1}.
 \end{example}
The hook $h=k-2$ is covered by the same method, once the set $\mathcal{D}$ is chosen so that its members remain distinct. The exponent $2k-3$ is not a sum of two exponents from $\{0, 1, \dots, k-1\}\setminus\{k-2\},$ so the product carrying it has to involve the twisted row.

\begin{theorem}\label{thm:: h = k-2}
    Let $\mathbb{F}_{q_0}$ be a subfield of $\mathbb{F}_q,$ and let $G$ be a subgroup of $\mathbb{F}_{q_0}^{*}$ of order $n$. For $b, c\in\mathbb{F}_{q_0}$ with $b\ne c$ and $G=\{1, \mu_1, \mu_2, \dots, \mu_{n-1}\},$ define $\alpha_{i}:=\frac{b-\mu_i c}{1-\mu_i},$ for $i\in [n-1]$. Suppose that $5\le k\le \frac{n-2}{2}$ and $h=k-2.$ If $\lambda\in \mathbb{F}_{q_0}^{*}\setminus G$ and $\eta\in \mathbb{F}_{q}\setminus\mathbb{F}_{q_0},$ then the code $\textnormal{RCTRS}_{k-2, 1}(\bm{\alpha}, b, c, \lambda, \eta)$ is a non-RS MDS code that is not equivalent to any CTRS code, extended CTRS code or TRS code with twist $t=1.$ The same holds for the extended code $\textnormal{RCTRS}_{k-2, 1}(\bm{\alpha}, b, c, \lambda, \eta, \infty)$.
\end{theorem}
\begin{proof}
    The code $\mathcal{C}:=\textnormal{RCTRS}_{k-2, 1}(\bm{\alpha}, b, c, \lambda, \eta)$ is MDS follows from Theorem \ref{thm:: MDS RCTRS code h and 1 construction subgroup}. With $h=k-2$ and the rows $\bm g_1, \dots, \bm g_k$ of the generator matrix labelled as before, put
    \begin{equation*}
        \begin{split}
            \mathcal{D}:=\{\bm{g}_1\star\bm{g}_1,\, \bm{g}_1\star\bm{g}_2,\, \dots,  \bm{g}_1\star\bm{g}_{k}\}\cup \{\bm{g}_2\star\bm{g}_{k},\,
    \bm{g}_3\star\bm{g}_{k},\dots,
    \bm{g}_{k}\star\bm{g}_{k}\}\\ \cup
    \{\bm{g}_{k-1}\star\bm{g}_{k-1},\, \bm{g}_{k-1}\star\bm{g}_{k-2}, \bm{g}_2\star\bm{g}_2\},
        \end{split}
    \end{equation*}
    which has $2k+2$ distinct elements. Denote its elements by $R_1, \dots, R_{2k+2}$ in the order displayed. For $1\le m\le 2k-1$ the polynomial part of $R_m$ is $x^{m-1},$ except for $m=k-1$ and $m=2k-2,$ where it is $x^{k-2}+\eta x^{k}$ and $x^{2k-3}+\eta x^{2k-1}$ respectively; the polynomial parts of $R_{2k}, R_{2k+1}$ and $R_{2k+2}$ are $x^{2k-4}+2\eta x^{2k-2}+\eta^{2}x^{2k},$ $x^{2k-5}+\eta x^{2k-3}$ and $x^{2},$ in that order. Apply the row operations
    $$R_{k-1}\rightarrow R_{k-1}-\eta R_{k+1},\qquad R_{2k+1}\rightarrow R_{2k+1}-R_{2k-4},$$
    $$R_{2k-2}\rightarrow \eta R_{2k-2}-R_{2k+1},\qquad R_{2k}\rightarrow R_{2k}-R_{2k-3}-2\eta R_{2k-1},\qquad R_{2k+2}\rightarrow R_{2k+2}-R_{3},$$
    in this order; the indices $k+1, 2k-4, 2k-3$ and $2k-1$ all lie in $\{1, \dots, 2k-1\}\setminus\{k-1, 2k-2\}$ because $k\ge 5,$ so the rows used carry the pure powers $x^{k}, x^{2k-5}, x^{2k-4}$ and $x^{2k-2}.$ After these operations the polynomial parts of $R_1, \dots, R_{2k+1}$ are non-zero scalar multiples of $1, x, \dots, x^{2k}$ in some order, the scalars being powers of $\eta,$ and $R_{2k+2}$ becomes $\lambda(b-c)^{2}\bm e_n.$ Choosing the $2k+1$ columns indexed by $\alpha_1, \dots, \alpha_{2k+1}$ together with the last column, which is possible since $n-1\ge 2k+1,$ the resulting $(2k+2)\times(2k+2)$ determinant is a non-zero scalar multiple of $\lambda(b-c)^{2}\underset{1\le i<j\le 2k+1}{\prod}(\alpha_j-\alpha_i),$ the scalar being $\pm$ a power of $\eta,$ and this is non-zero. Hence $\dim(\mathcal{C}^{\star 2})\ge 2k+2,$ and the inequivalence follows exactly as in Theorem \ref{thm: non-RS TRS for any h}. 
    
    Finally, write $\mathcal{C}^{\infty}$ for the extended code. Deleting its last coordinate returns $\mathcal{C},$ so Lemma \ref{lem:: puncturing} gives $\dim((\mathcal{C}^{\infty})^{\star 2})\ge \dim(\mathcal{C}^{\star 2})\ge 2k+2.$ The code $\mathcal{C}^{\infty}$ is MDS by Theorem \ref{thm:: MDS RCTRS code h and 1 construction subgroup}, and the families it is compared with have Schur squares of dimension at most $2k+1,$ so the same conclusion holds for $\mathcal{C}^{\infty}$.\hfill$\square$
\end{proof}

Table \ref{tab:: comparison} collects the outcome of the preceding three sections. Each row records a family of codes of length $n$ and dimension $k$ against which our codes are compared, together with the dimension of its Schur square; each of the last three columns records one of the three regimes of the hook and names the result which places the corresponding RCTRS codes outside that family. The dimensions in the second column are supplied by Lemma \ref{lem:: schur square of GRS} for Reed--Solomon codes, by \cite[Theorems 1 and 2]{liu2025column} together with Lemma \ref{cor:: schur square bounds}(a) for column-twisted codes, and by Lemma \ref{cor:: schur square bounds}(b), (c) for twisted codes with $t=1$.

\begin{table}[htbp]
\centering
\renewcommand{\arraystretch}{1.3}
\setlength{\tabcolsep}{6pt}
\begin{tabular}{@{}llccc@{}}
\toprule
Family of $[n,k]$ codes & & \multicolumn{3}{c}{$\textnormal{RCTRS}_{h, 1}(\bm{\alpha}, b, c, \lambda, \eta)$ and its extended code}\\
\cmidrule(l){3-5}
&
 & $h=0$ & $h=k-1$ & $1\le h\le k-2$\\
 & $\dim \mathcal{C}^{\star 2}$ & $(\ge 2k+1)$ & $(\ge 2k+1)$ & $(\ge 2k+2)$\\
\midrule
RS and extended RS & $2k-1$
 & Thm.\ \ref{thm:: MDS code 0 and 1 construction subgroup}
 & Thm.\ \ref{thm:: MDS code k-1 and 1 construction subgroup}
 & Thms.\ \ref{thm: non-RS TRS for any h}, \ref{thm: not CTRS TRS for any h=1}, \ref{thm:: h = k-2}\\
$\textnormal{CTRS}(\bm{\alpha}', b', c', \lambda')$ & $2k$
 & Thm.\ \ref{thm:: non-equivalence of MDS code 0 and 1 construction subgroup and CTRS}
 & Thm.\ \ref{thm:: non-equivalence k-1 and 1 and CTRS}
 & Thms.\ \ref{thm: non-RS TRS for any h}, \ref{thm: not CTRS TRS for any h=1}, \ref{thm:: h = k-2}\\
$\textnormal{CTRS}(\bm{\alpha}', b', c', \lambda', \infty)$ & $2k$
 & Thm.\ \ref{thm:: non-equivalence of MDS code 0 and 1 construction subgroup and CTRS}
 & Thm.\ \ref{thm:: non-equivalence k-1 and 1 and CTRS}
 & Thms.\ \ref{thm: non-RS TRS for any h}, \ref{thm: not CTRS TRS for any h=1}, \ref{thm:: h = k-2}\\
$\textnormal{TRS}_{k, 0, 1}(\bm{\beta}, \eta')$ & $\le 2k$
 & Thm.\ \ref{thm:: non-equivalence of MDS code 0 and 1 construction subgroup and CTRS}
 &---
 & Thms.\ \ref{thm: non-RS TRS for any h}, \ref{thm: not CTRS TRS for any h=1}, \ref{thm:: h = k-2}\\
$\textnormal{TRS}_{k,  h', 1}(\bm{\beta}, \eta'),\ h'\ne 0$ & $\le 2k+1$
 & ---
 & ---
 & Thms.\ \ref{thm: non-RS TRS for any h}, \ref{thm: not CTRS TRS for any h=1}, \ref{thm:: h = k-2}\\
\bottomrule
\end{tabular}
\caption{Schur square dimension comparisons of RCTRS codes and their extended versions with RS, CTRS, and their extended versions, as well as TRS codes. Since equivalent codes have Schur squares of the same dimension, the strict inequalities yield the corresponding inequivalence results.}
\label{tab:: comparison}
\end{table}

\section{Conclusion}\label{Section 7}
    In this article, we introduced a new class of codes, called row-column twisted Reed--Solomon (RCTRS) codes, motivated by the works of Beelen et al. \cite{beelen2017twisted} and Liu et al. \cite{liu2025column}. We first presented generator matrices for RCTRS and extended RCTRS codes. We then established necessary and sufficient conditions for these codes to be MDS and provided explicit constructions satisfying these conditions. By analyzing the dimensions of their Schur squares, we showed that several subfamilies of these MDS codes are equivalent to neither Reed--Solomon codes, column-twisted Reed--Solomon codes, nor a single twisted Reed--Solomon code with twist $t=1$. Thus, our construction yields new families of MDS codes that are inequivalent to these previously known classes, thereby providing a novel class of non-RS MDS codes.

    For future work, it would be natural to derive explicit parity-check matrices for RCTRS codes and investigate efficient decoding algorithms for these codes. Another direction is to explore broader generalizations involving twists with $t\ne 1$, as well as multiple row and column twists. Finally, studying the potential applications of RCTRS codes in code-based cryptography, particularly in view of their distinct Schur-square dimensions, may provide further insights into their cryptographic potential.
    
\section*{Declarations}
\subsection*{Conflict of Interest}
All authors declare that they have no conflict of interest.

\section*{Acknowledgments}
The first author acknowledges the support of the Council of Scientific and Industrial Research (CSIR) India, under grant no. 09/0086(13310)/2022-EMR-I.

\bibliographystyle{abbrv}
\bibliography{subfield}

\end{document}